\documentclass[preprint,12pt,authoryear]{elsarticle}

\usepackage{amssymb}
\usepackage{amsmath}

\usepackage{hyperref}
\usepackage{float}
\usepackage{subfig}
\usepackage{caption}
\usepackage{graphicx}
\usepackage{rotating}
\usepackage{tablefootnote}
\usepackage{multirow}
\usepackage{tabularx}
\usepackage{tabularray}
\usepackage{adjustbox}
\usepackage[english]{babel} 
\usepackage[autostyle]{csquotes} 

\usepackage{natbib}
\setcitestyle{numbers}
\setcitestyle{square}
\usepackage{xcolor}
\usepackage{ulem}
\usepackage{soul}
\usepackage{tabu}
\usepackage{threeparttable}

\journal{} 

\begin{document}

\begin{frontmatter}

\title{Underwater Image Enhancement by Convolutional Spiking Neural Networks}

\author[1]{Vidya Sudevan\corref{cor1}} 
\ead{100060499@ku.ac.ae} 
\author[1,2]{Fakhreddine Zayer} 
\ead{fakhreddine.zayer@ku.ac.ae} 
\author[1]{Rizwana Kausar} 
\ead{100062832@ku.ac.ae}
\author[1,3]{Sajid Javed} 
\ead{sajid.javed@ku.ac.ae} 
\author[1,4]{Hamad Karki} 
\ead{hamad.karki@ku.ac.ae} 
\author[1,5]{Giulia De Masi} 
\ead{giulia.demasi@sorbonne.ae} 
\author[1,2]{Jorge Dias} 
\ead{jorge.dias@ku.ac.ae} 

\cortext[cor1]{Corresponding author}

\affiliation[1]{organization={Center for Autonomous Robotic Systems},
            addressline={Khalifa University}, 
            country={UAE}}
\affiliation[2]{organization={Department of Computer and Information Engineering},
            addressline={Khalifa University},
            country={UAE}}
\affiliation[3]{organization={Department of Computer Science},
            addressline={Khalifa University},
            country={UAE}}
\affiliation[4]{organization={Department of Mechanical and Nuclear Engineering},
            addressline={Khalifa University},
            country={UAE}}
\affiliation[5]{organization={Department of Science and Engineering},
            addressline={Sorbonne University Abu Dhabi},
            country={UAE}}

\begin{abstract}
Underwater image enhancement (UIE) is fundamental for marine applications, including autonomous vision-based navigation. Deep learning methods using convolutional neural networks (CNN) and vision transformers advanced UIE performance. Recently, spiking neural networks (SNN) have gained attention for their lightweight design, energy efficiency, and scalability. This paper introduces UIE-SNN, the first SNN-based UIE algorithm to improve visibility of underwater images. UIE-SNN is a 19- layered convolutional spiking encoder-decoder framework with skip connections, directly trained using surrogate gradient-based backpropagation through time (BPTT) strategy. We explore and validate the influence of training datasets on energy reduction, a unique advantage of UIE-SNN architecture, in contrast to the conventional learning-based architectures, where energy consumption is model-dependent. UIE-SNN optimizes the loss function in latent space representation to reconstruct clear underwater images. Our algorithm performs on par with its non-spiking counterpart methods in terms of PSNR and structural similarity index (SSIM) at reduced timesteps ($T=5$) and energy consumption of $85\%$. The algorithm is trained on two publicly available benchmark datasets, UIEB and EUVP, and tested on unseen images from UIEB, EUVP, LSUI, U45, and our custom UIE dataset. The UIE-SNN algorithm achieves PSNR of \(17.7801~dB\) and SSIM of \(0.7454\) on UIEB, and PSNR of \(23.1725~dB\) and SSIM of \(0.7890\) on EUVP. UIE-SNN achieves this algorithmic performance with fewer operators (\(147.49\) GSOPs) and energy (\(0.1327~J\)) compared to its non-spiking counterpart (GFLOPs = \(218.88\) and Energy=\(1.0068~J\)). Compared with existing SOTA UIE methods, UIE-SNN achieves an average of \(6.5\times\) improvement in energy efficiency. The source code is available at \href{https://github.com/vidya-rejul/UIE-SNN.git}{UIE-SNN}.
\end{abstract}



\begin{keyword}
Energy efficient architecture, Lightweight neural networks, Spiking neural networks, Underwater Image Enhancement.
\end{keyword}

\end{frontmatter}


\section{Introduction}
Underwater image enhancement (UIE) remains a critical challenge in computer vision \cite{zhou2023underwater}. UIE is vital for various applications, including marine monitoring \cite{goodwin2022unlocking}, vision-based navigation \cite{leordeanu2021driven}, aquatic biodiversity assessment \cite{odilov2024utilizing}, fisheries management, and marine species counting \cite{ovalle2022use}. However, UIE development is hindered by challenges such as poor lighting, color distortion, and water turbidity, all of which degrade image quality and compromise the performance of computer vision techniques. Several methods have been proposed to tackle UIE challenges. Ancuti et al. \cite{ancuti2012enhancing} combined contrast-enhanced and color-corrected images using multi-scale fusion with four weights to highlight important pixels. Li et al. \cite{li2018emerging} introduced a weakly supervised color correction technique that aligns underwater images with open-air counterparts through cross-domain mapping. GAN-based methods, which balance a generator and a discriminator, have shown promise over CNN techniques. For example, Li et al. \cite{li2017watergan} developed WaterGAN, a two-stage network using synthetic underwater images for training. Park et al. \cite{park2019adaptive} introduced a CycleGAN with adaptive weights to enhance images, correct color, and extract features. Despite their advantages, GAN architectures require careful convergence for optimal performance. Vision Transformers (ViTs) have recently achieved remarkable results in computer vision tasks like object detection \cite{liu2022two}, recognition \cite{si2023token}, and visual tracking \cite{hao2022umotma}, and have also impacted UIE. Guo et al. \cite{guo2023underwater} introduced URanker, a conv-attentional image Transformer using histogram tokens to address global degradation and dynamic cross-scale correspondence for local degradation. Another approach by Guo et al. \cite{ren2022reinforced} integrated the Swin Transformer into U-Net, creating URSCT-SESR, which combines convolutions with attention mechanisms to enhance local and global attention. While these deep learning models have advanced UIE performance, they are resource-intensive and computationally demanding, limiting their applicability on edge and mobile platforms where power efficiency is critical. 

The light absorption and scattering by water molecules and suspended particles significantly alter the perceived colors and reduce the contrast of images. These distortions vary spatially and temporally, complicating the task of developing robust enhancement techniques. Such conditions demand innovative solutions that are not only effective in enhancing visual quality but are also computationally efficient to accommodate the energy constraints in underwater operations. Traditional approaches, while capable, often fall short in energy efficiency, necessitating the exploration of novel computational paradigms.

Spiking Neural Networks (SNNs) have recently gained attention for various computer vision tasks, including object detection \cite{kim2020spiking}, tracking \cite{xiang2024spiking}, neuromorphic vision \cite{osswald2017spiking}, pose estimation \cite{kreiser2018pose}, structural health monitoring \cite{pang2020case}, image recognition \cite{ngu2022effective}, and image classification \cite{iakymchuk2015simplified}. Known as \enquote{third generation neural networks}, SNNs emulate neuronal computation, incorporating a temporal dimension through precise spike timing \cite{maass1997networks}. This allows them to efficiently process spatio-temporal data with significantly reduced energy consumption \cite{roy2019towards}, making them well-suited for resource-constrained environments like underwater settings. SNNs also leverage temporal binary codes, replacing multiply-accumulate (MAC) operations with energy-efficient accumulation (ACC) operations \cite{castagnetti2023spiden}. Their asynchronous and parallel nature aligns well with neuromorphic hardware, such as IBM's TrueNorth \cite{merolla2014million} and Intel's Loihi \cite{davies2018loihi}, further boosting computational speed and energy efficiency.

Efficient deep SNNs for image processing are developed through ANN-to-SNN conversion or direct SNN training \cite{li2024deep}. Conversion transforms pre-trained ANNs into SNNs using techniques like weight normalization and threshold balancing for image classification \cite{diehl2015fast}, and clamped, quantized training to minimize conversion loss \cite{yan2021near}. Object detection has used signed neurons with imbalanced thresholds to convert Tiny YOLO into an SNN \cite{kim2018spiking}. Direct SNN training, whether supervised or unsupervised, includes spike-timing-dependent plasticity (STDP) for classification \cite{diehl2015unsupervised} and recognition \cite{liu2020unsupervised}. Supervised methods, despite challenges with non-differentiable spiking neuron functions, employ techniques like smoothing spikes and spatial-temporal backpropagation (STBP) with surrogate gradients \cite{wu2018spatio}. Hybrid methods combining STDP and STBP are also explored \cite{chakraborty2021fully}. However, these often demand significant memory, leading to extensions like semantic segmentation \cite{kim2022beyond} and ASF-BP for efficient gradient calculation \cite{wu2021training}. Surrogate gradient methods dominate SNN research, using spiking functions forward and continuous approximations backward \cite{kim2021revisiting}, with local learning \cite{kaiser2020synaptic} and three-factor rules \cite{kusmierz2017learning} enhancing accuracy. Techniques like BPTT \cite{kim2021revisiting}, spike-based BP focusing on LIF neurons \cite{lee2020enabling}, and spike response kernel convolution \cite{shrestha2018slayer} are common. SNNs typically convert real-valued inputs into discrete spikes using rate coding, through Poisson coding, despite being popular \cite{han2020rmp,lee2020enabling}, may introduce randomness. The direct coding \cite{liao2023convolutional} is a simpler method where the initial layer directly processes real-valued data into spike sequences. It enhances accuracy and reduces simulation time, and is used in the proposed UIE-SNN algorithm.

In traditional deep learning, residual connections, as exemplified by VGG models, improve performance by preventing accuracy degradation as depth increases \cite{he2016deep,vicente2022keys}. These connections reduce optimization complexity which can lead to accuracy saturation. Trainable spiking ResNet models have been developed for SNNs, allowing deeper networks to be trained with performance comparable to VGGs \cite{zheng2021going,lee2020enabling}. The proposed UIE-SNN algorithm adopts a spike-output-to-spike-output (S2S) connection approach in its skip-connected layers. Recent studies have investigated SNNs for pixel-wise prediction tasks, achieving promising results in image reconstruction and denoising, primarily for low-resolution images and Gaussian noise removal \cite{roy2019synthesizing, comcsa2021spiking, castagnetti2023spiden, li2024deep, cao2024spiking, song2024learning}. However, these methods fall short for high-resolution underwater images with non-linear haze. The potential of SNNs in underwater computer vision, particularly for UIE, is largely untapped. Unlike traditional deep learning, SNNs are lightweight, energy-efficient, and capable of real-time implementation on neuromorphic hardware. 

Despite their potential, there is currently no existing literature detailing the development and implementation of SNN-based algorithms for processing real-world underwater images with visibility degradations. This knowledge gap motivated the development of the current study, which provides a comprehensive exploration of SNN-based UIE methods. This work introduces a novel SNN architecture for UIE, dubbed as UIE-SNN, a spiking convolutional encoder-decoder model with skip connections designed to enhance underwater image visibility. Specifically, this article focuses on carefully selecting critical hyperparameters, such as threshold membrane potential, timesteps, and architectural depth, to optimize real-world image processing while leveraging the energy efficiency of SNNs. By addressing this unexplored area, the study aims to advance the use of SNNs for energy-efficient underwater image enhancement in practical and constrained environments.

\begin{figure}[!h]
    \centering
    \subfloat[]{\includegraphics[width=0.8\textwidth]{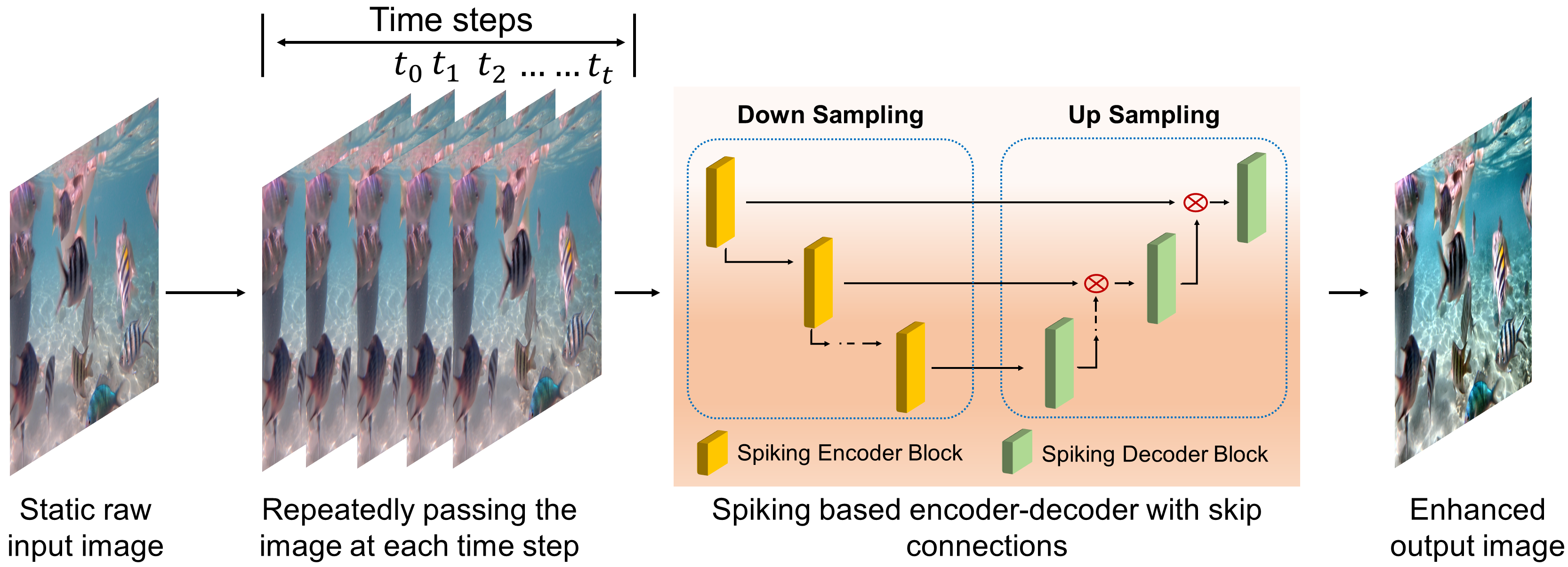}
    \label{fig:intro_overview}}
    \vspace{-0.3cm}
    \subfloat[]{\includegraphics[width=0.8\textwidth]{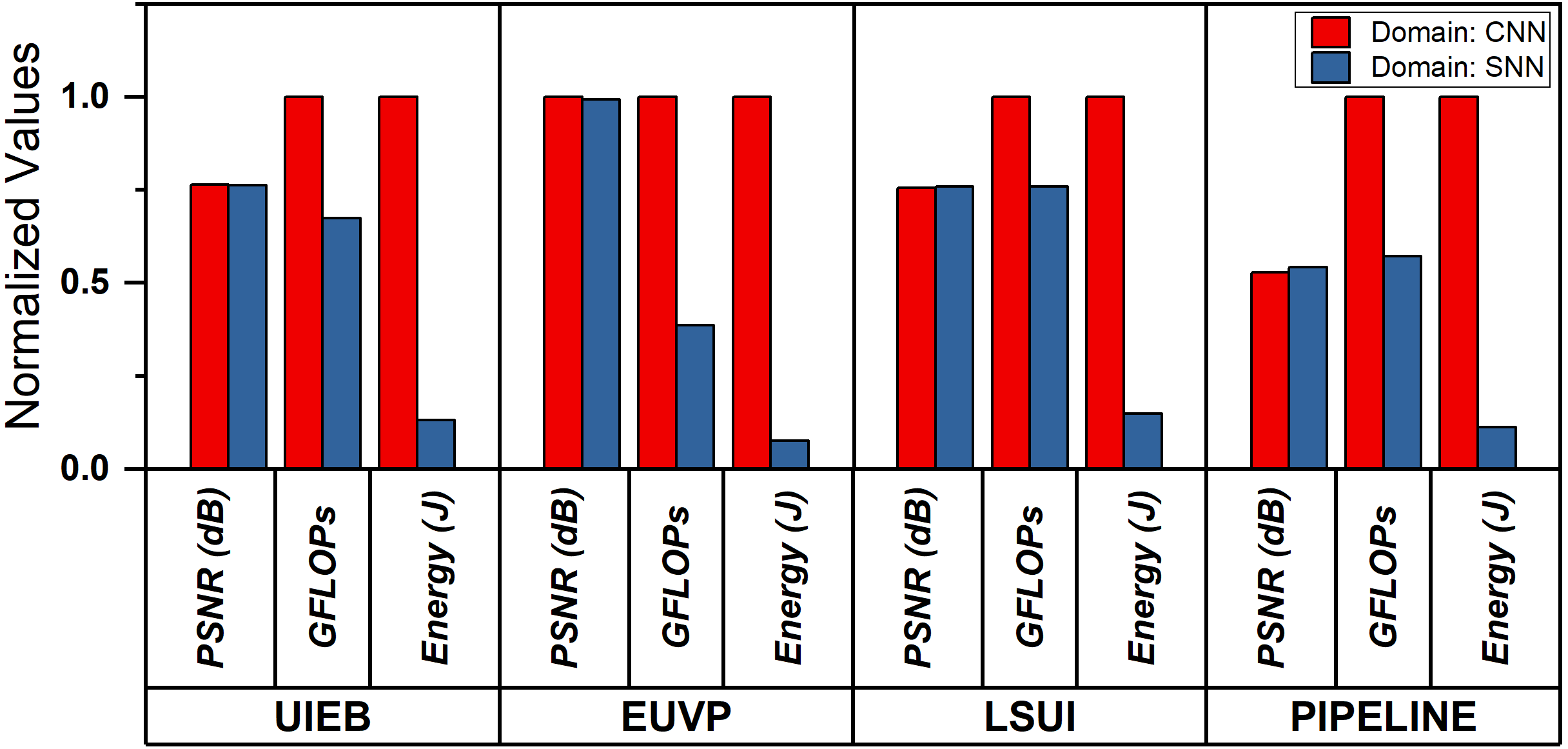}
    \label{fig:intro_compare}}
    \vspace{-0.3cm}
    \caption{(a) Overview of UIE-SNN framework: The raw images are first converted into time-dependent image sequences. These sequences are then fed to the convolutional spiking-based encoder-decoder structure to extract high-level features and thereby reconstruct the desired visibility-enhanced images by minimizing the artifacts., and (b) Evaluation of UIE-SNN with its CNN counterpart: The proposed UIE-SNN framework is demonstrating comparable algorithmic performance at significantly reduced GFLOPs and energy consumption with its CNN counterpart.}
    \label{fig:intro_fig}
\end{figure}

In our approach, the raw underwater images are transformed into time-dependent sequences by repeatedly processing static images over a set number of timesteps. These sequences are fed into the first convolutional layer of the encoder block, followed by a leaky integrate-and-fire (LIF) neuron layer, which converts continuous pixel values into sparse spike representations. Subsequent layers apply convolution operations to these spikes, utilizing spike-output-to-spike-output (S2S) skip connections at each encoder-decoder layer. In the decoder block, the membrane potential from the final LIF neuron layer is recorded at the last timestep and compared to the pixel values of a reference image, allowing the computation of Mean Squared Error (MSE) for backpropagation during training. UIE-SNN employs a surrogate gradient-based approach with a fast-sigmoid function for backpropagation through time (BPTT) network training. Our model capitalizes on the inherent advantages of SNNs, such as low energy consumption and rapid processing, to enhance UIE tasks. The high-level representation of the proposed UIE-SNN framework and its performance comparison with a non-spiking counterpart are illustrated in Fig. \ref{fig:intro_overview} and Fig. \ref{fig:intro_compare}. 

This work's novelty lies in the direct training of the SNN framework using membrane potentials over reduced timesteps, a strategy not previously explored in SNN-based UIE tasks. Key contributions include: 
\begin{itemize}
    \item \textbf{First SNN-based Algorithm for Underwater Image Enhancement:}  We present UIE-SNN, a pioneering approach by utilizing SNN for UIE tasks, addressing the lack of SNN implementations for real-world UIE tasks. The framework significantly reduces energy consumption (by approximately $85\%$) compared to non-spiking counterpart and optimizes synaptic operations through direct training based on output neuron membrane potentials.
    \item \textbf{Direct Training Approach:}  A direct training \cite{henkes2024spiking} method that utilizes membrane potential dynamics and a surrogate gradient-based BPTT learning rule is proposed here. This approach enhances pixel prediction accuracy by treating the UIE task as a regression problem.
    \item \textbf{Energy Efficiency in SNN Frameworks:} This research highlights the potential of SNN frameworks to reduce energy consumption when trained on complex, realistic data representations. The model's performance is validated on unseen underwater images, demonstrating its robustness across various datasets.
    \item \textbf{Domain-Specific Parameter Analysis:} We conduct a comprehensive analysis of the impact of hyperparameters on image enhancement, particularly focusing on threshold and timestep selection. The study provides both qualitative and quantitative assessments on domain-specific datasets like UIEB and EUVP, offering valuable insights into energy-efficient underwater image processing.
\end{itemize}

\section{Related Works}
In this section, the existing works on image processing tasks using \(2^{nd}\) and \(3^{nd}\) generation ANNs are presented. The UIE methods using CNNs, GAN, attention modules, and transformer blocks are detailed in the image processing methods using \(2^{nd}\) generation ANNs. Since there are no \(3^{rd}\) generation ANN-based methods available for UIE tasks, the methods referring to pixel-wise prediction tasks are detailed. 
\subsection{Image Processing using \(2^{nd}\) Generation ANNs}

Li et al. \cite{li2020underwater} introduced WaterNet, a simple, fully CNN-based model for UIE. To predict the confidence maps, this model processes three derived inputs along with the original input. Before fusion, three Feature Transformation Units (FTUs) are applied to the inputs to reduce color casts and artifacts introduced by the WB, HE, and GC algorithms. The refined inputs are then multiplied by the confidence maps to generate the final enhanced image. Shallow-UWnet \cite{naik2021shallow} uses shallow ConvBlocks consisting of Conv-ReLU-Dropout layers and skip connections, and is trained with a multi-term loss function, including MSE and VGG perceptual loss. Fu et al. \cite{fu2022uncertainty} presented a probabilistic network (PUIE-Net) to learn the enhancement distribution of degraded underwater images, combining a conditional variational autoencoder with adaptive instance normalization, followed by a consensus process to predict a deterministic result from the samples. With the rise of GAN models, WaterGAN \cite{li2017watergan} was developed to generate realistic underwater images from in-air images and depth pairings in an unsupervised pipeline, aimed at color correction for monocular underwater images. The pipeline includes a depth estimation network that reconstructs a coarse relative depth map and a color restoration network based on a fully convolutional encoder-decoder architecture, SegNet. PUGAN \cite{cong2023pugan} is a physical model-guided GAN for UIE, featuring a parameters estimation subnetwork that learns physical model inversion parameters, using generated color enhancement images as auxiliary information for a two-stream interaction enhancement sub-network. Dual discriminators enforce style and content constraints, improving authenticity and visual quality. Finally, DGD-cGAN \cite{gonzalez2024dgd} targets haze and color cast removal induced by water, restoring true underwater scene colors. This model uses two generators: the first predicts the dewatered scene, while the second learns the underwater image formation process, employing a custom loss function based on transmission and veiling light components. 

Ucolor \cite{li2021underwater} learns feature representations from various color spaces and emphasizes the most discriminative features using a channel-attention module. It incorporates domain knowledge by utilizing the reverse medium transmission map as attention weights. LANet \cite{liu2022adaptive}, a supervised adaptive attention network for UIE, begins with a multiscale fusion module that combines different spatial information. It then employs a parallel attention module to focus on illuminated features and significant color information, using both pixel and channel attention. An adaptive learning module retains shallow information and adaptively learns important features. The network is trained using a multinomial loss function combining mean absolute error and perceptual loss, with an asynchronous training mode enhancing performance. Deep-WaveNet \cite{sharma2023wavelength} is optimized with traditional pixel-wise and feature-based cost functions. It utilizes a distinctive receptive field size for each image channel, driven by its wavelength, which helps in learning diverse local and global features. These features are adaptively refined using a block attention mechanism. DICAM \cite{tolie2024dicam} addresses proportional degradations and non-uniform color cast by using inception modules over each color channel to extract feature maps at three scales. The extracted feature maps are weighted with a CAM to capture the importance of different degradations. The combined feature maps are then refined with CAM to enhance the image's color richness. U-Trans \cite{peng2023u} incorporates a channel-wise multi-scale feature fusion transformer module and a spatial-wise global feature modeling transformer module, enhancing the network's focus on color channels and spatial areas with more severe attenuation. UIE-Convformer \cite{wang2024uie} is a hybrid method combining CNNs and a feature fusion Transformer. It employs a multi-scale U-Net structure for extracting rich texture and semantic information, while the feature fusion transformer module handles global information fusion. A jump fusion connection module between the encoder and decoder fuses multi-scale features through bidirectional cross-connection and weighted fusion, enriching the feature information for image reconstruction. Transformer-based models demonstrate performance comparable to CNNs due to their ability to construct long-distance dependencies and extend receptive fields across various visual tasks. However, their application and deployment are challenged by the large number of parameters involved.

\subsection{Image Processing using \(3^{rd}\) Generation ANNs}
Roy et al. \cite{roy2019synthesizing} introduced a method to synthesize images from multiple modalities within a spike-based representation. They utilized a spiking autoencoder to encode inputs into compact spatio-temporal representations, which were then decoded for image reconstruction. The training process involved computing the membrane potential loss at the output layer and backpropagating it using a sigmoid approximation of the neuron's activation function to ensure differentiability. Later, in 2021, a spiking autoencoder with temporal coding and pulse-based training was presented \cite{comcsa2021spiking}, demonstrating the importance of inhibition for memorizing inputs over extended periods before generating the expected output spikes. Castagnetti et al. \cite{castagnetti2023spiden} developed the first SNN-based Gaussian image denoising model, using surrogate gradient learning and backpropagation through time (BPTT) directly in the spike domain. In 2024, two advancements were made: a spiking U-Net using multi-threshold neurons was proposed, leveraging an ANN-SNN conversion and fine-tuning pipeline based on pre-trained U-Net models \cite{li2024deep}, and a training-free approach was developed using threshold guidance for SNN-based diffusion models in Gaussian image denoising \cite{cao2024spiking}. Additionally, ESDNet \cite{song2024learning}, an image-deraining model, introduced a spiking residual block that converted inputs into spike signals, optimizing membrane potentials adaptively with attention weights to mitigate information loss from discrete binary activations. While these methods have shown promise in image reconstruction and denoising, they are limited to low-resolution images and Gaussian noise removal, making them inadequate for high-resolution, complex underwater image sequences with non-linear haze distributions.

\section{Convolutional Spiking-Based Encoder-Decoder Framework for visibility enhancement.}
\begin{figure}[!ht]
\includegraphics[width=\textwidth]{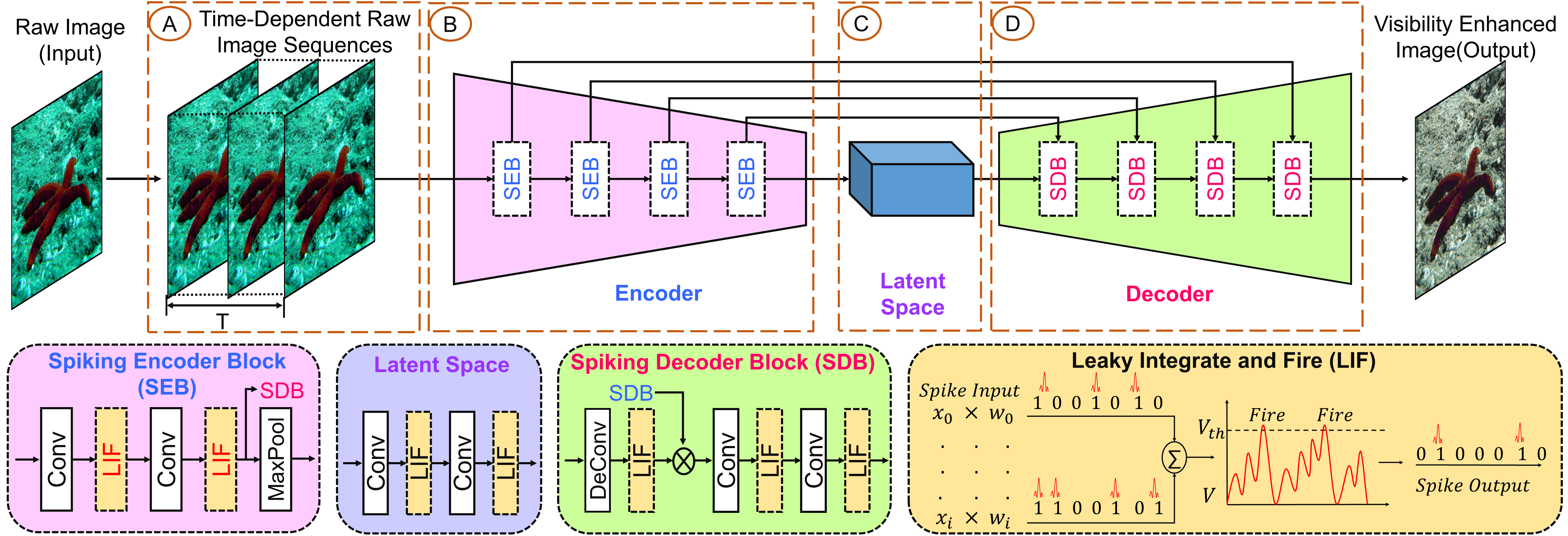}
\centering
\caption{Detailed UIE-SNN Model: In phase (A), raw images are converted into time-dependent sequences. Phase (B) involves the first CL and LIF neuron, transforming pixel values into sparse spike representations while extracting contextual features in the encoder block. Phase (C) shows the latent space, and in phase (D), the visibility-enhanced image is reconstructed by minimizing reconstruction loss.}
\label{fig:arch_overview}
\end{figure}

Fig. \ref{fig:arch_overview} illustrates the overview of the proposed UIE-SNN designed for enhancing the visibility of underwater images. The design consists of a sequential arrangement of convolutional layers (CLs) with LIF neurons, creating a symmetrical encoder-decoder structure to efficiently extract and reconstruct the visibility-enhanced images. Each Spiking Encoder Block (SEB) consists of two convolutional layer instances and a LIF neuron module, followed by a max pooling layer. The max pooling layers downsample the feature maps to achieve a reduced feature representation. 

The latent space utilizes two CLs with LIF neurons to reduce the feature representation. During the decoder phase, the feature maps are upsampled using transposed CLs, and the spatial information is restored by LIF neurons. Each spiking decoder block consists of two CLs along with corresponding LIF modules. Finally, the enhanced image is reconstructed by utilizing the membrane potential of the last LIF neuron layer. The overall architecture is distinguished by skip connections, which enable the flow of information between encoder and decoder layers to maintain spatial features and minimize information loss. The feature maps produced by the encoder and the upsampled feature maps produced by the decoder are concatenated using S2S connections, which improves the extraction of spatially rich information from the visually degraded underwater images.

\subsection{Spiking Neuron Model}\label{sec:neuron_model}
\begin{figure}[!h]
\includegraphics[width=0.6\textwidth]{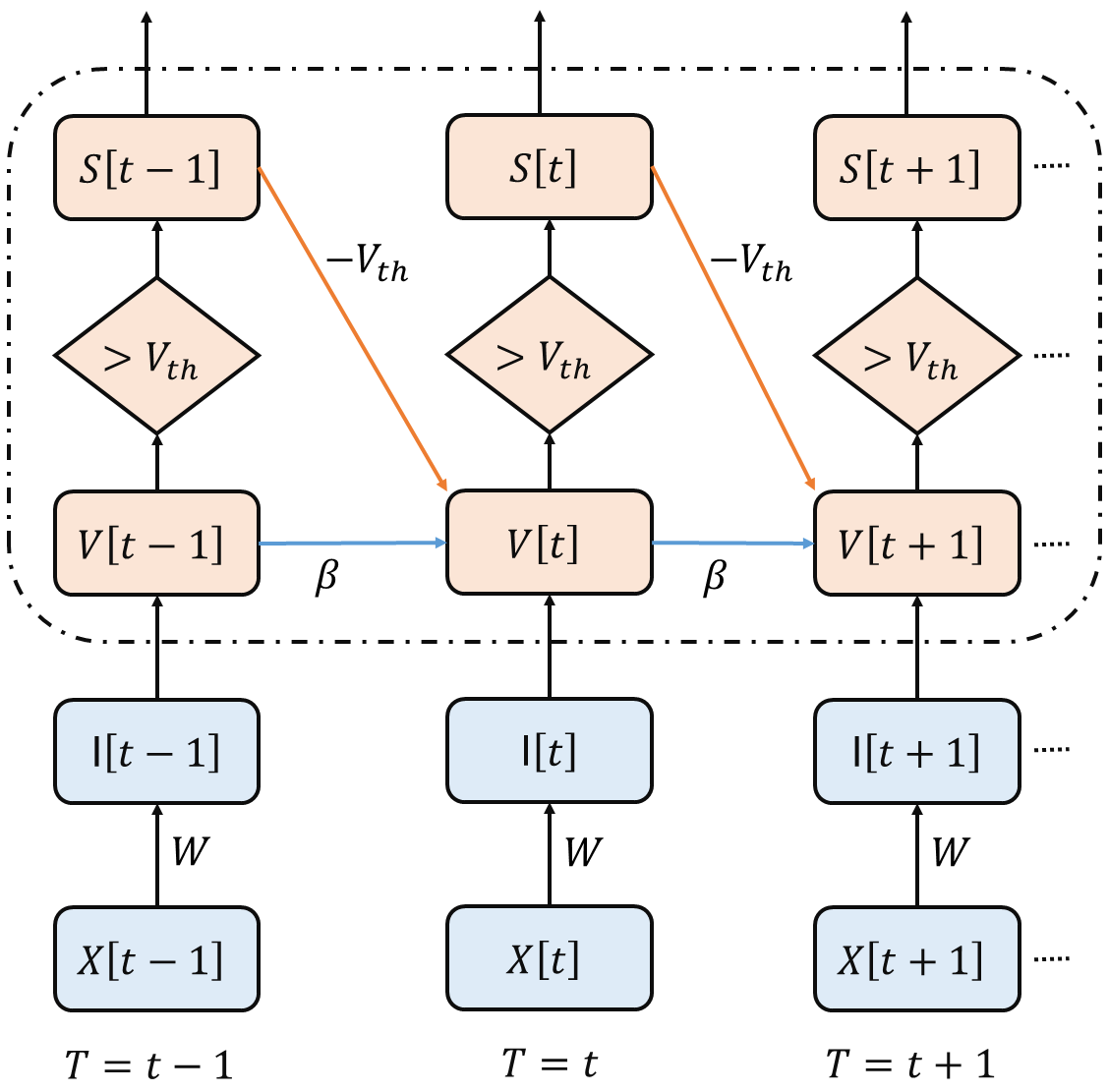}
\centering
\caption{Computational graph of LIF neuron over the timesteps.}
\label{fig:lif_computation}
\end{figure}

The computation graph of a spiking neuron driven by input neurons via plastic synapses are simulated using a biologically plausible threshold-based Leaky-Integrate-and-Fire (LIF) model is presented in Fig.\ref{fig:lif_computation}. The LIF neuron dynamics is modeled using a RC circuit \cite{eshraghian2023training} and is represented as:

\begin{equation}
    \centering
    \tau \frac{d V(t)}{dt}=-V(t)+I(t)R
    \label{eq:lif_dynamics}
\end{equation}
where, \(\tau\) is the time constant (\(\tau=RC\)), \(R\) and \(C\) are the resistance and capacitance of the RC circuit respectively. \(I(t)\) is the input current. 

The approximate solution to (\ref{eq:lif_dynamics}) described as in (\ref{eq:approximate_soln}) is calculated using the Euler method to ensure compatibility with sequence-based neural network. 

\begin{equation}
    \centering
    V[t]=\beta V[t-1]+(1-\beta)I[t]
    \label{eq:approximate_soln}
\end{equation}
where, \(\beta = e^{-1/\tau}\), is the decay rate of \(V[t]\).

In the learning-based frameworks, the weighting factor \((1- \beta)\) of the input \(I[t]\) is considered as a learnable parameter. For a simplified representation, the input \(I[t]\) is considered as a single neuron input \(X[t]\). Therefore,

\begin{equation}
    \centering
    V[t]=\beta V[t-1]+W X[t]-S[t-1]V_{th}
\end{equation}
where, \(W\) is the learned synaptic weight, \(V_{th}\) is the threshold membrane potential, \(S[t-1]\in \left\{ 1,0 \right\}\) is the spike output generated by the neuron. The neuron firing is described as \(S[t]=\Theta (V[t]-V_{th})\). \(\Theta (V[t]-V_{th})\) is the Heaviside function defined as:

\begin{equation}
    \Theta (V[t]-V_{th}) = 
    \begin{cases}
    1, & V[t] \ge V_{th} \\
    0, & otherwise
    \end{cases}
\end{equation}

\subsection{Encoding of Input Images}

\begin{figure}[!h]
\includegraphics[width=0.95\textwidth]{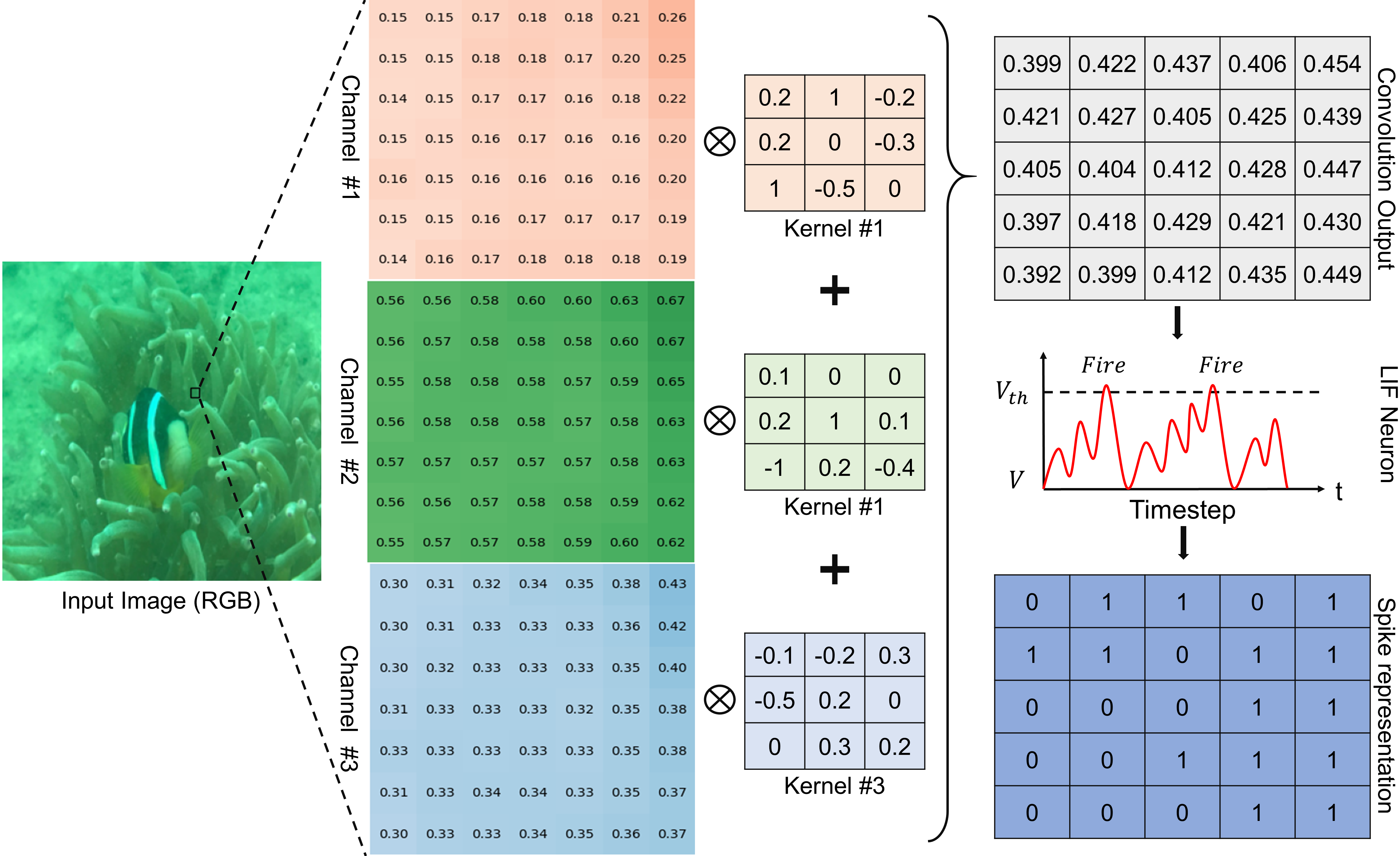}
\centering
\caption{Overview of spike encoding: In the first stage, feature maps are generated when each channel of the input RGB images undergoes convolution with a respective kernel. These convolved data is then fed into spiking neuron, where the membrane potential (\(V\)) is updated over the timesteps. When \(V\) exceeds the threshold membrane potential (\(V_{th}\)), a spike is fired, and \(V\) resets to a predefined value.}
\label{fig:spike_coding}
\end{figure}

The first layer of the architecture transforms the three-channel raw underwater images into spike representations. The raw images serve as continuous input to the architecture for a user-defined timestep duration. Following the first convolutional layer, the spiking neurons convert the incoming presynaptic inputs into their equivalent binary spikes. The overview of the spike coding process during a single timestep, where the continuous-valued input signal is converted into its spike-based representation is presented in Fig.\ref{fig:spike_coding}.

The spike-to-image conversion occurs in the final layer. The incoming spikes are fed to the final convolutional layer followed by a LIF neuron. The membrane potential of the LIF neurons at the last timestep, without resetting, is directly considered as the corresponding pixel-wise values. 

\subsection{UIE-SNN: Architectural Description}

\begin{figure}[!h]
\includegraphics[width=\textwidth]{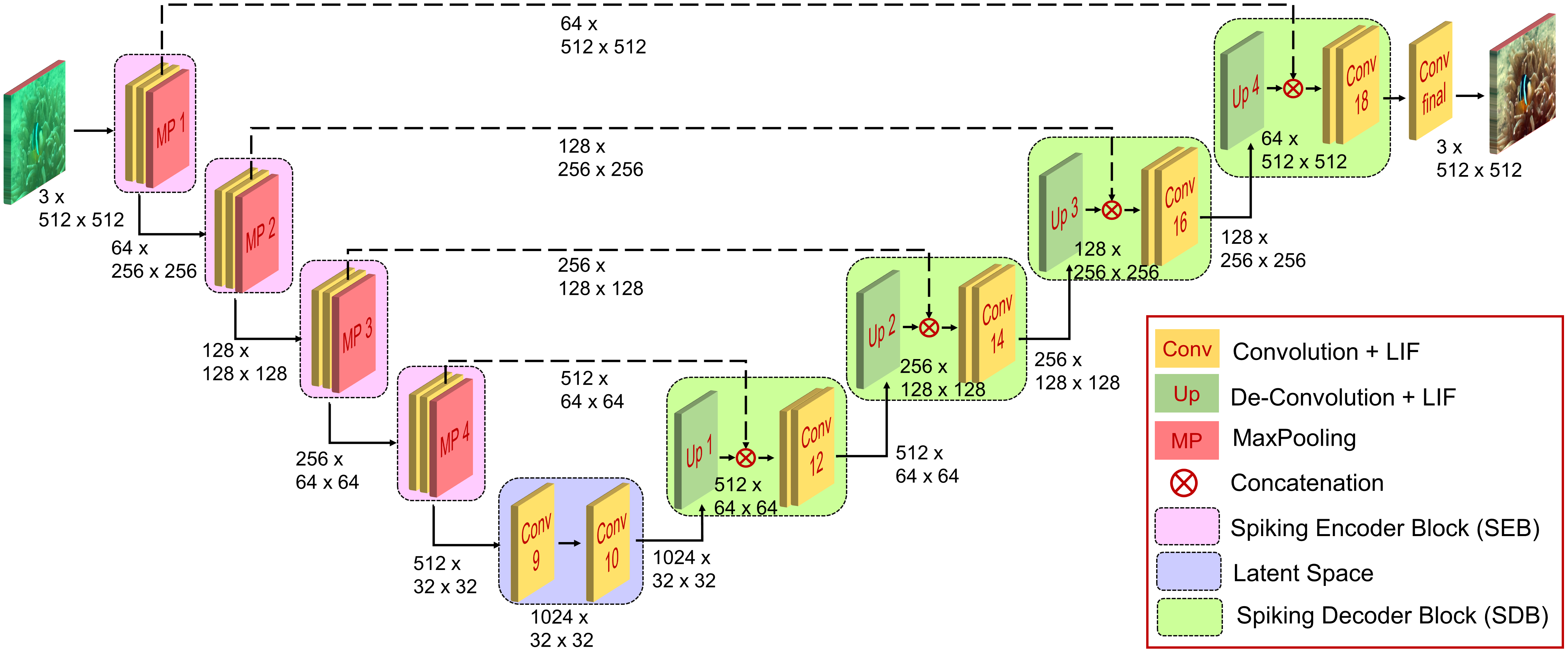}
\centering
\caption{Detailed structural description of UIE-SNN for a single timestep: At a single timestep, the continuous-valued input image is first converted into its equivalent spike representation at the first convolution layer foloowed by the LIF neuron. The encoder path consists of four SEB blocks to extract the high-level spatial and temporal information. At the latent space representation, additional high-level features are extracted using convolutional spiking blocks.In the decoding phase, SDB blocks progressively up-sample the feature maps, integrating skip connections from the encoder to preserve spatial features, and finally the visibility enhanced image with the same spatial resolution of the input image is reconstructed using the final convolutional spiking layer. The size of output feature map at the end of each layer is presented.}
\label{fig:uie-snn_arch_details}
\end{figure}

The UIE-SNN model is a 19-layered convolutional spiking encoder-decoder framework. It follows a U-net \cite{ronneberger2015u} inspired architecture with spiking neurons, as shown in Fig.\ref{fig:uie-snn_arch_details}. 

It consists of an encoder, a latent space representation, and a decoder. The encoder consists of four SEB blocks, each includes a 2D convolutional layers followed by a LIF neuron and a max pooling layer for downsampling. Specifically, each stage includes two convolutional layers with \(3\times3\) kernels, where the number of channels progressively doubles from 64 to 512, while the spatial resolution is halved. The outputs of these layers (skip connections) are saved for concatenation during decoding. The bottleneck further applies two convolutional layers with 1024 channels to extract latent features. 

The decoder mirrors the encoder structure with four stages of upsampling using deconvolution layers, followed by concatenation with corresponding skip connections and LIF neuron activations. Finally, the output convolutional layer with a \(3\times3\) kernel reconstructs the visibility enhanced image. The input and output dimensions are \(3\times512\times512\), and the total number of network parameters are 31.03 million. A fixed threshold is used for the LIF neurons. The threshold and timestep selection criteria are detailed in section \ref{sec:ablation}. 

\subsection{Network Training using BPTT} \label{bptt_algo_revise}

The UIE-SNN model is trained using the surrogate gradient-based BPTT strategy to update the weights of the network, ensuring loss minimization. In reference to Fig.\ref{fig:lif_computation}, the backpropagation algorithm in SNN-based model at a single timestep is defined as:

\begin{equation}
    \frac{\partial \mathcal{L}}{\partial W}=\frac{\partial \mathcal{L}}{\partial S}\frac{\partial S}{\partial V}\frac{\partial V}{\partial I}\frac{\partial I}{\partial W}
    \label{eq:BP_SNN}
\end{equation}
The output spike generated during the forward pass is given as:
\begin{equation}
    S[t]= \Theta (V[t]-V_{th}) = 
    \begin{cases}
    1, & V \ge V_{th} \\
    0, & otherwise
    \end{cases}
    \label{eq:forward}
\end{equation}
The derivative calculated during the backward pass for (\ref{eq:forward}) is:
\begin{equation}
    \frac{\partial S}{\partial V}= \delta (V-V_{th}) = 
    \begin{cases}
    +\infty , & V = V_{th} \\
    0, & otherwise
    \end{cases}
\end{equation}
where, \(\delta(.)\) is the Dirac-Delta function.

This implies that the term \(\frac{\partial S}{\partial V}\) will almost always be zero, thereby inhibiting the learning process. In this case, the network training will be unstable with the use of \(\delta(.)\) function to calculate the gradient and apply the gradient descent.

The surrogate gradient-based method with fast-sigmoid function is used in this work to mitigate the dead neuron problem \cite{eshraghian2023training}. The threshold-centered membrane potential is defined as \(\tilde{V} = V-V_{th}\). The fast-sigmoid function and its derivative used for backpropagation is represented as:

\begin{equation}
    \tilde{S} \approx \frac{\tilde{V}}{(1+\lambda\left| \tilde{V} \right|)}
\end{equation}

\begin{equation}
    \frac{\partial \tilde{S}}{\partial V} \approx \frac{1}{(1+\lambda\left| \tilde{V} \right|)^{2}}
\end{equation}
where, \(\lambda\) modulates the smoothness of the surrogate function.

Since SNNs operates in the temporal domain, the weight \(W\) influences the membrane potential and loss across all time steps. Since the weights are shared across all timesteps (\(W[0]=W[1]= ...=W[T]=W\)), the global gradient of the loss \(\mathcal{L}\) with respect to the weight \(W\) over all past timesteps \( p\le t\) is defined as:

\begin{equation}
    \frac{\partial \mathcal{L}}{\partial W}=\sum_{t=1}^{T}\frac{\partial \mathcal{L}[t]}{\partial W}=\sum_{t=1}^{T}\sum_{ p\le t}^{}\frac{\partial \mathcal{L}[t]}{\partial W[p]}\frac{\partial  W[p]}{\partial W}= \sum_{t=1}^{T}\sum_{ p\le t}^{}\frac{\partial \mathcal{L}[t]}{\partial W[p]}
\end{equation}

For instance, if only the immediate prior influence corresponding to a timestep \(p=t-1\) is taken into account, the backward pass needs to trace back in time by a single step. In this case, the influence of \(W[t-1]\) on \(\mathcal{L}[t]\) is defined as:

\begin{equation}
    \frac{\partial \mathcal{L}[t]}{\partial W[t-1]}=\frac{\partial \mathcal{L}[t]}{\partial S[t]}\frac{\partial \tilde{S}[t]}{\partial V[t]}\frac{\partial V[t]}{\partial V[t-1]}\frac{\partial V[t-1]}{\partial I[t-1]}\frac{\partial I[t-1]}{\partial W[t-1]}
\end{equation}
where \(\frac{\partial V[t]}{\partial V[t-1]}\) is the decay value defined for a LIF neuron; \(\frac{\partial V[t-1]}{\partial I[t-1]} = 1\), and \(\frac{\partial I[t-1]}{\partial W[t-1]}\) is the presynaptic input.

By combining surrogate gradients with BPTT, the training of SNNs becomes both feasible and efficient. The surrogate gradient approximations enable non-zero gradients even when the membrane potential is near the threshold. The BPTT strategy ensures that the temporal dependencies are captured across timesteps. The surrogate gradient used in this work aligns with established methods in the literature \cite{eshraghian2023training}, and provides smooth gradient flow while preserving the temporal dynamics essential for SNNs.

\subsection{Loss Function}
During the training process, the MSE loss metric is used to reduce the reconstruction error between the reference image and the network generated image from the latent space representation of the raw input images. The MSE error is fed back to the architecture and is used to update the weights and biases. 

\begin{equation}
    \centering
    \mathcal{L}_{MSE}=\frac{1}{l\times w}\sum_{i=1}^{l}\sum_{j=1}^{w}\left ( J_{i,j}-\hat{J}_{i,j} \right )^{2}
\end{equation}
where, \(l \times w\) is the length and width of the image, \(J\) is the reference image, \(\hat{J}\) is the reconstructed image, and \((i,j)\) are the corresponding membrane potential.

\section{Results and Analysis}
In this article, we introduce a spiking convolution framework trained directly using a surrogate gradient-based backpropagation strategy to improve the visibility of underwater image sequences. This approach allows us to achieve comparable performance with reduced timesteps and lower energy consumption. The following sections detail the experiments conducted, including dataset descriptions, training parameter settings, and various ablation studies. 

\begin{figure}[!ht]
\includegraphics[width=\textwidth]{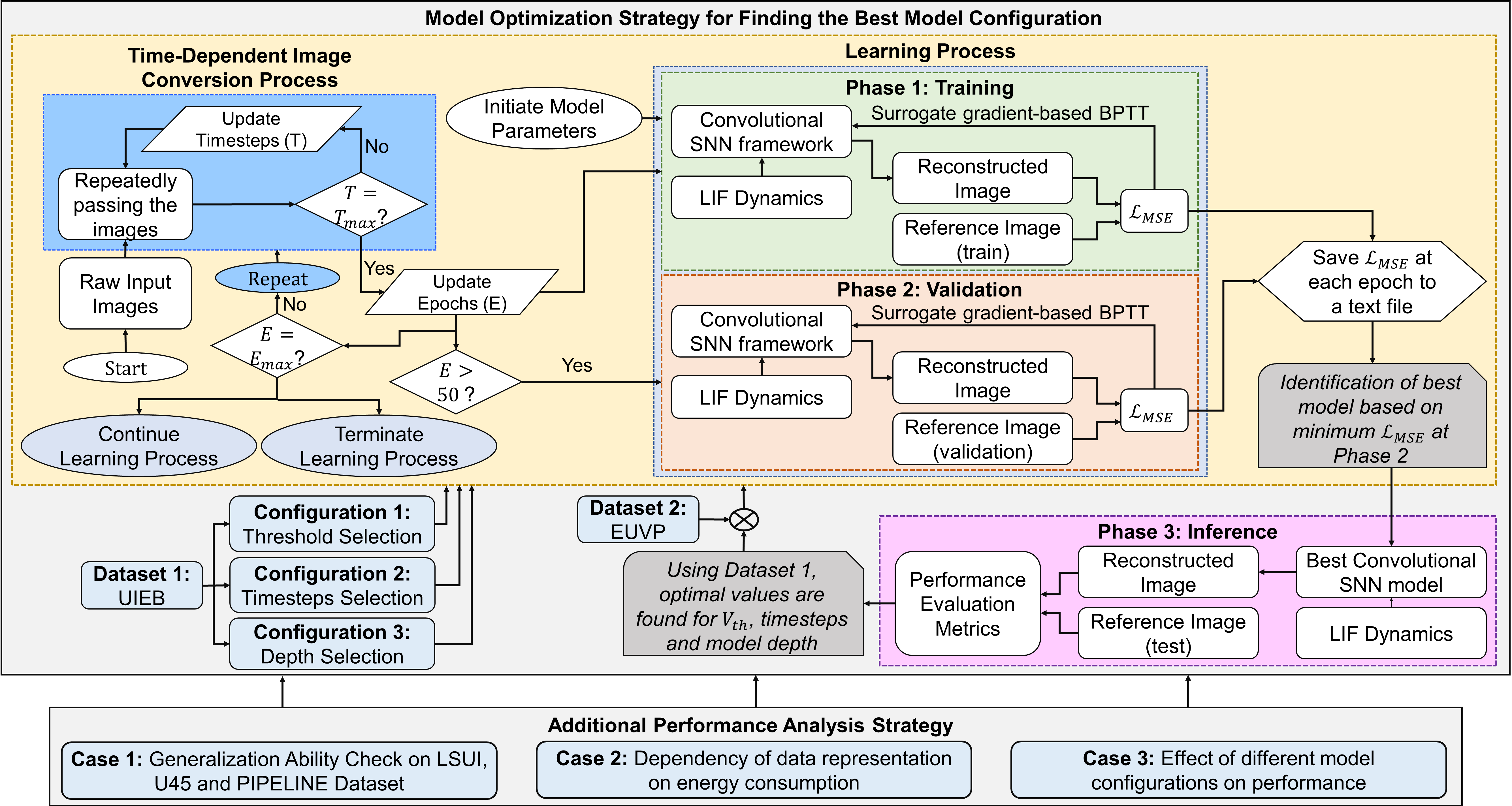}
\centering
\caption{Training, validation, and parameter optimization strategy adopted for the implementation of the proposed UIE-SNN framework.}
\label{fig:methodology}
\end{figure}

The proposed framework is implemented using snnTorch \cite{eshraghian2023training} and PyTorch library, and is trained on a high-performance NVIDIA A100 GPU with 80GB RAM. All models in this study were trained for 200 epochs to ensure convergence, with the validation starts at the \(50^{th}\) epoch to allow the model a warm-up period. The best-performing model was selected based on the lowest validation loss throughout the epochs. We used the Adam \cite{kingma2014adam} optimizer with a learning rate of 0.001 to ensure stable convergence and efficient gradient updates. During the training and inference process, all the images are resized to \(512 \times 512\). The selection of neuron thresholds, required timesteps, and the depth of the encoder-decoder layer was based on empirical analysis. The parameter selection criteria and the performance evaluation of the proposed UIE-SNN framework, compared to its non-spiking counterpart, are presented in the following subsections. For computing the percentage of energy reduction, the non-spiking counterpart with the same depth of the encoder-decoder layer was used as the baseline for comparison with the SNN-based frameworks. The overview of the methodology adopted in this paper is presented in Fig.\ref{fig:methodology}.

\subsection{Dataset Descriptions}
The proposed UIE-SNN framework is trained separately on UIEB \cite{li2020underwater}, a real-underwater dataset and EUVP \cite{islam2020fast}, a synthetic underwater dataset and cross-validated on UIEB, EUVP, LSUI \cite{peng2023u}, U45 \cite{li2019fusion}, and a custom pipeline dataset \cite{sudevan2022multisensor}.

\begin{enumerate}
\item \textbf{UIEB}, a real-underwater dataset consists of \(890\) paired images with corresponding reference images and \(60\) unpaired images. The reference images were synthetically generated and selected using a combination of conventional image enhancement techniques.
\item \textbf{EUVP}, a large-scale synthetically generated underwater dataset containing \(12K\) paired and \(8K\) unpaired images generated using CycleGAN-based model. The reference images were obtained through a human-robot cooperative experiment and from publicly available underwater videos. 
\item \textbf{LSUI}, a real-underwater dataset containing \(4279\) images, each paired with a synthetically generated clear image via a reference comparison method. Semantic segmentation map and a medium transmission map is also been provided with each underwater image pair.
\item \textbf{U45}, a real-world underwater non-reference dataset consisting of 45 images, divided equally among three categories: green, blue, and haze-like underwater scenes.
\item \textbf{PIPELINE}, a custom, synthetically generated underwater pipeline dataset consisting of 300 image pairs, each captured at three different turbidity levels. The reference images were directly obtained during a pipeline tracking experiment conducted in an indoor pool setup.
\end{enumerate}

\subsection{Evaluation metrics}
The following evaluation metrics have been used to effectively assess the proposed UIE-SNN framework's algorithmic and hardware-oriented performance.

\subsubsection{Algorithmic Performance Evaluation Metrics} \label{sec:algo_metrics}
For a quantitative assessment of the proposed UIE-SNN framework, we have used both the full-reference (PSNR and SSIM) and the non-reference (BRISQUE, UCIQE, AND UIQM) evaluation metrics. 
\paragraph{\textbf{Peak Signal-to-Noise Ratio (PSNR)}} It measures the quality of visibility-enhanced images by comparing them to the original, calculating the ratio of signal power to the noise affecting the image \cite{huynh2008scope}. A higher PSNR indicates that the enhanced image closely resembles the original. 

\begin{equation}
    \centering
    PSNR(J,\hat{J})=20log_{10}\left ( \frac{P_{max}}{MSE(J,\hat{J})} \right )
\end{equation}
where, \(P_{max}\) is the maximum possible pixel value.

\paragraph{\textbf{Structural Similarity Index (SSIM)}} SSIM evaluates the similarity between two images by comparing structural information, luminance, and contrast \cite{wang2004image}. Unlike PSNR, which focuses on pixel differences, SSIM provides a value between -1 and 1, where 1 indicates perfect similarity. 

\begin{equation}
    \centering
    SSIM\left ( J,\hat{J} \right )=\frac{\left ( 2\mu_{J}\mu_{\hat{J}} + C_{1} \right )\left ( 2\sigma _{J\hat{J} } + C_{2} \right )}{\left ( \mu _{J}^{2} +\mu _{\hat{J}}^{2} + C_{1} \right )\left ( \sigma _{J}^{2} +\sigma _{\hat{J}}^{2} + C_{2} \right )}
\end{equation}
where \(\mu_{J}\) and \(\mu_{\hat{J}}\) are the means, \(\sigma _{J}^{2}\) and \(\sigma _{\hat{J}}^{2}\) are the variances, \(\sigma _{J\hat{J} }\) is the covariance. \(C_{1}\) and \(C_{2}\) are constants.

\paragraph{\textbf{Blind/Referenceless Image Spatial Quality Evaluator (BRISQUE)}} BRISQUE \cite{mittal2012no} is a no-reference metric that assesses image quality based on naturalness, using features derived from luminance coefficients. The underlying assumption is that high-quality images exhibit certain statistical regularities. BRISQUE uses a machine learning model trained on a large dataset of images with known quality scores to predict the image quality without a reference image, with lower scores indicating better quality.

\paragraph{\textbf{Underwater Color Image Quality Evaluation (UCIQE)}} UCIQE \cite{yang2015underwater} assesses the effectiveness of UIE algorithms for underwater images by measuring chroma variance, saturation, and contrast, which together reflect the colorfulness and sharpness of the image. 
\begin{equation}
    \centering
    UCIE = \left (U_{1} \times \sigma_{c} \right ) + \left (U_{2} \times con_{l} \right ) + \left (U_{3} \times \mu_{s} \right )    
\end{equation}
where \(\sigma_{c}\) is chroma standard deviation, \(con_{l}\) is luminance contrast, \(\mu_{s}\) is average saturation, and \(U_{1}\), \(U_{2}\), and \(U_{3}\) are the weighted coefficients.

\paragraph{\textbf{Underwater Image Quality Measure (UIQM)}} UIQM \cite{panetta2015human}  combines measures of colorfulness (UICM), sharpness (UISM), and contrast (UIConM) into a single metric. The weights for each components (\(M_{1}\), \(M_{2}\), and \(M_{3}\)) are adjusted based on the application, with higher UIQM values indicating better image quality.

\begin{equation}
    \centering
    \begin{split}
    UIQM & = \left ( M_{1} \times UICM \right ) + \left ( M_{2} \times UISM \right ) \\
         & + \left ( M_{3} \times UIConM  \right )
\end{split}
\end{equation}

\subsubsection{Energy Measurement}
The energy consumption for CNNs and SNNs are computed based on the total number of Floating Point Operations (FLOPs) and  Synaptic Operators (SOPs) respectively. The energy consumption is based on the standard CMOS technology \cite{horowitz20141} as shown in Table  \ref{table:energy_cmos}. 

\begin{table}[H]
\renewcommand{\arraystretch}{1.3} 
\caption{Energy consumption for 45nm CMOS process.}
\label{table:energy_cmos}
\centering
\begin{tabular}{l|c}
\hline
\bfseries Floating Point (FP) Operation & \bfseries Energy (pJ) \\ 
\hline \hline
32 bit FP MULT                         & 3.7                  \\ 
32 bit FP ADD                          & 0.9                  \\ 
32 bit FP MAC (\(E_{MAC}\))            & 4.6                  \\ 
32 bit FP ACC (\(E_{ACC}\))              & 0.9                  \\ 
\hline
\end{tabular}
\end{table}

The FLOPs and SOPs for each convolution layer \(l\) in CNN and SNN are computed as follows:

\begin{equation}
    \centering
    \begin{matrix}
    CNN: & FLOPs_{CNN}(l)=l_{Cin} \times l_{Cout} \times l_{k}^{2} \times l_{l} \times l_{w}\\
    & \\
    SNN: & SOPs_{SNN}(l)=FLOPs_{CNN}(l) \times S_{r}(l)
\end{matrix}
\end{equation}
where, \(l_{Cin}\) is the number of input channels at layer \(l\), \(l_{Cout}\) is the number of output channels at layer \(l\), \(l_{k}\) is the size of the kernel at layer \(l\), \(l_{l}\) is the length of the output feature map at layer \(l\), \(l_{w}\) is the width of the output feature map at layer \(l\) and \(S_{r}(l)\) is the spike rate \cite{kim2022beyond} at each SNN layer \(l\).

\begin{equation}
    \centering
    \begin{matrix}
    S_{r}(l) & =\frac{Number\;of\;spikes\;over\;all\;timesteps}{Number\;of\;neuron}     \\
     & \\
     & = \frac{\sum_{c=1}^{l_{Cout}}\sum_{i=1}^{l_{l}}\sum_{j=1}^{l_{w}}\sum_{t=1}^{T}S^{t}_{l}(i,j,c)}{l_{Cout}\times l_{l} \times l_{w}}
    \end{matrix}
\end{equation}

The spike rate \(S_{r}(l)\) effectively scales the computational cost of SNNs by accounting for spike-driven sparse computations. Since SNNs operate on binary spikes rather than continuous activations, the overall number of operations can be significantly reduced compared to traditional CNNs. 

Given the operation counts per layer, the total inference energy for CNNs and SNNs across all layers can be determined as follows:
\begin{equation}
    \centering
    \begin{matrix}
CNN: & E_{CNN}=\sum_{l=1}^{L}FLOPs_{CNN}(l) \times E_{MAC}\\ 
 & \\ 
SNN: & E_{SNN}=\sum_{l=1}^{L}SOPs_{SNN}(l) \times E_{ACC}
\end{matrix}
\end{equation}
where \(L\) is the total number of layers in the architecture, \(E_{MAC}\) represents the energy per MAC operation, and \(E_{ACC}\) represents the energy per accumulation operation. \(E_{MAC}\) and  \(E_{ACC}\) can be obtained from Table  \ref{table:energy_cmos}.

The percentage of energy reduction is calculated as:

\begin{equation}
    \centering
    \Delta E (\%) =\frac{E_{CNN}-E_{SNN}}{E_{CNN}} \times 100
\end{equation}

\subsection{Performance Evaluation} \label{perform_eval}
Since there are no SNN-based UIE methods available in the literature, this section focuses on the performance evaluation of the proposed UIE-SNN algorithm compared to its non-spiking (CNN-based) counterpart. The algorithms are trained separately using the UIEB and EUVP datasets. The algorithmic and hardware-oriented metric evaluation is tested and cross-validated using other datasets. 

Additionally, the effect of training data representations on energy consumption is presented in section \ref{sec:data_energy_effect}. The selection process for the threshold membrane potential, timesteps, and architectural depth is described in section \ref{sec:ablation}. All results presented in this section compare the UIE-SNN (indicated as SNN) with its non-spiking counterpart (indicated as CNN). 

The detailed comparison of the proposed UIE-SNN algorithm with existing SOTA SNN-based methods for pixel-wise prediction tasks is presented in Table \ref{table:snn_in_imgproc}. Comparison with existing analytical and learning-based SOTA UIE methods are presented in Table \ref{table:sota_conv} and Table \ref{table:sota_learn} respectively.

Based on the empirical studies conducted, the threshold is selected as \(0.25\), the number of timesteps \(T\) is set to \(5\), and the depth of the encoder-decoder layer is set to \(4\) for all subsequent evaluations. The algorithmic performance evaluation of the proposed model is conducted using both full-reference and non-reference evaluation metrics described in section \ref{sec:algo_metrics}. Hardware-oriented performance is assessed using inference rate, GFLOPs, energy consumption, and \(\Delta E (\%)\), utilizing the UIEB and EUVP datasets, as shown in Table \ref{table:performance_quantity}. The algorithmic performance is evaluated for both the SNN and its non-spiking counterpart. Since the computation of GFLOPs and energy consumption in the CNN domain is independent of spike rates across the layers, these values remain constant at \(218.88 ~GFLOPs\) and \(1.0068 ~J\), respectively, for a layer depth of four.

\begin{table}[!ht]
\centering
\caption{Performance evaluation of the proposed UIE-SNN (SNN-based) framework with its non-spiking (CNN-based) counterpart. The evaluation is based on algorithmic and hardware-oriented performance metrics comparison of both frameworks trained using UIEB and EUVP datasets and cross-validated on other datasets.}
\label{table:performance_quantity}
\renewcommand{\arraystretch}{1.3} 
\adjustbox{width=\textwidth}{\begin{tabular}{c|c|cc|cc|cc|cc|cc|c|c|c}
\hline
\multicolumn{2}{c|}{\textbf{Dataset}} & \multicolumn{10}{c|}{\textbf{Algorithmic Performance}} & \multicolumn{3}{c}{\textbf{Hardware-Oriented Metrics}} \\
\hline
\multirow{2}{*}{\textbf{Train}} & \multirow{2}{*}{\textbf{Test}} & 
\multicolumn{2}{c|}{\textbf{PSNR (dB) ($\uparrow$)}} & 
\multicolumn{2}{c|}{\textbf{SSIM ($\uparrow$)}} & 
\multicolumn{2}{c|}{\textbf{BRISQUE ($\downarrow$)}} & 
\multicolumn{2}{c|}{\textbf{UCIQE ($\uparrow$)}} & 
\multicolumn{2}{c|}{\textbf{UIQM ($\uparrow$)}} & 
\textbf{GSOPs} & 
\textbf{Energy} & 
\textbf{$\Delta E$ (\%)}\\
& & \textbf{\textit{SNN}} & \textbf{\textit{CNN}} & \textbf{\textit{SNN}} & \textbf{\textit{CNN}} & \textbf{\textit{SNN}} & \textbf{\textit{CNN}} & \textbf{\textit{SNN}} & \textbf{\textit{CNN}} & \textbf{\textit{SNN}} & \textbf{\textit{CNN}} & ($\downarrow$) & (J) ($\downarrow$) & ($\uparrow$) \\
\hline \hline
\multirow{5}{*}{\textbf{UIEB}} & \textbf{UIEB} & 17.7801 & 17.8122 & 0.7454 & 0.8205 & 35.8177 & 30.5761 & 0.7906 & 0.8721 & 2.2557 & 2.9549 & 147.49 & 0.1327 & \textbf{86.82} \\
& \textbf{EUVP} & 20.2711 & 20.3193 & 0.7438 & 0.8085 & 42.5508 & 52.6769 & 0.8827 & 1.0197 & 3.0046 & 2.9034 & 86.03 & 0.0774 & 92.31 \\
& \textbf{LSUI} & 18.7367 & 18.4830 & 0.7725 & 0.8391 & 46.2530 & 45.7331 & 0.6599 & 0.6339 & 3.2314 & 3.0814 & 165.21 & 0.1487 & \textbf{85.23} \\
& \textbf{U45} & - & - & - & - & 36.0409 & 46.2819 & 0.6871 & 0.6949 & 2.1795 & 2.8572 & 66.82 & 0.0601 & \textbf{94.03} \\
& \textbf{PIPELINE} & 14.2353 & 14.6828 & 0.6220 & 0.7482 & 42.1659 & 46.6570 & 0.5011 & 0.4575 & 2.8305 & 2.2349 & 87.49 & 0.0787 & \textbf{92.18} \\
\hline
\multirow{5}{*}{\textbf{EUVP}} & \textbf{UIEB} & 16.6787 & 16.8379 & 0.6943 & 0.7822 & 27.1973 & 40.1282 & 1.1038 & 1.2116 & 2.2075 & 2.7059 & 170.91 & 0.1538 & 84.72 \\
& \textbf{EUVP} & 23.1725 & 23.3290 & 0.7890 & 0.8641 & 39.3322 & 55.3430 & 1.3882 & 1.5687 & 2.9047 & 2.7544 & 84.65 & 0.0762 & \textbf{92.43} \\
& \textbf{LSUI} & 17.6992 & 17.6309 & 0.7501 & 0.8290 & 38.8840 & 47.4674 & 0.7273 & 0.7382 & 3.0483 & 2.8188 & 166.03 & 0.1494 & 85.16 \\
& \textbf{U45} & - & - & - & - & 44.3889 & 49.9663 & 0.8174 & 0.7667 & 2.4981 & 2.5577 & 69.85 & 0.0629 & 93.76 \\
& \textbf{PIPELINE} & 12.6490 & 12.3310 & 0.6958 & 0.8060 & 41.2632 & 54.1544 & 0.5208 & 0.5134 & 2.8000 & 1.7400 & 125.23 & 0.1127 & 88.81 \\
\hline
\end{tabular}}
\end{table}

\begin{figure}[!h]
\includegraphics[width=\textwidth]{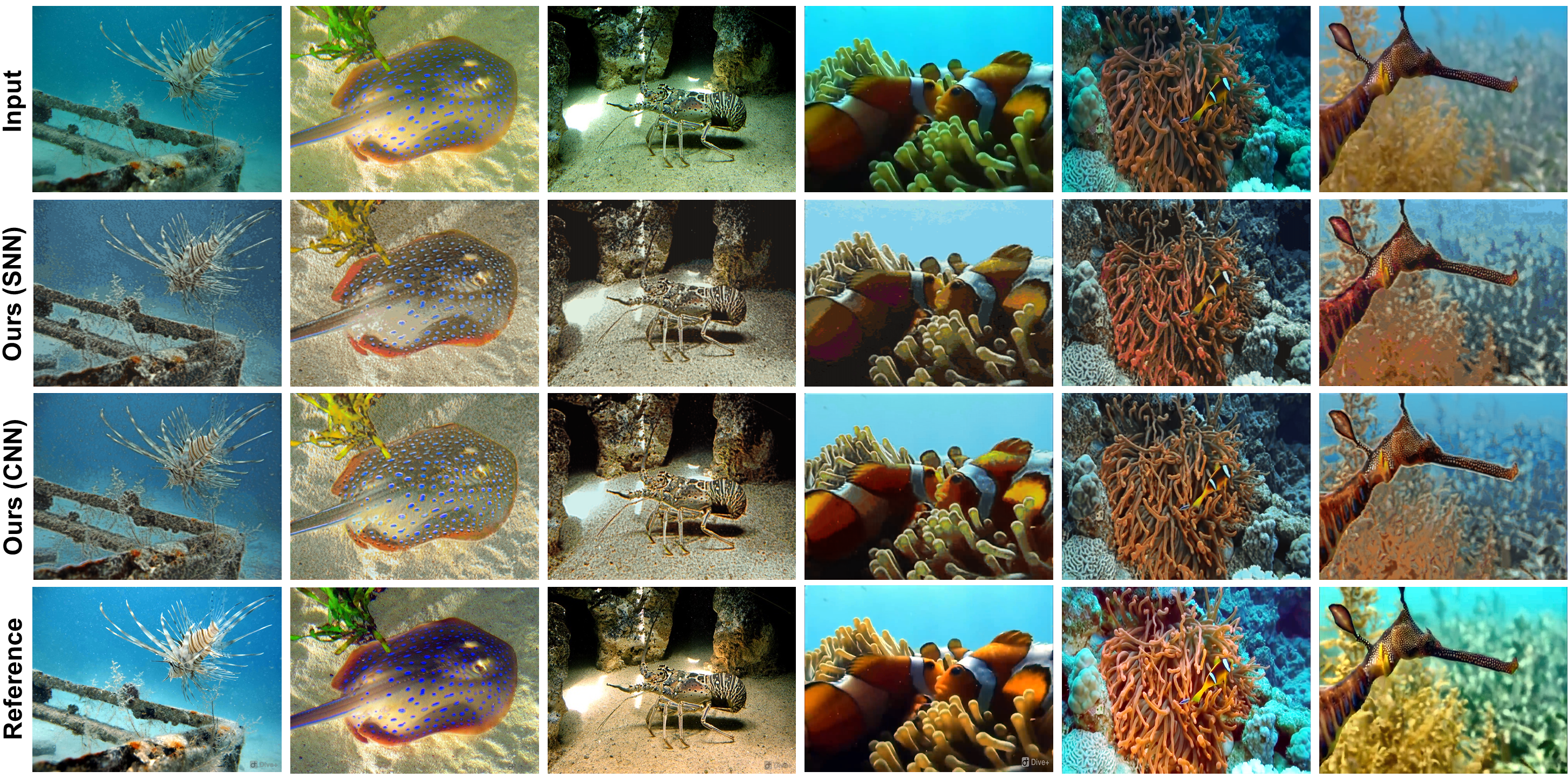}
\centering
\caption{Qualitative analysis of proposed UIE-SNN framework for visibility enhancement on UIEB test samples.}
\label{fig:uieb_eval}
\end{figure}

\begin{figure}[!h]
\includegraphics[width=\textwidth]{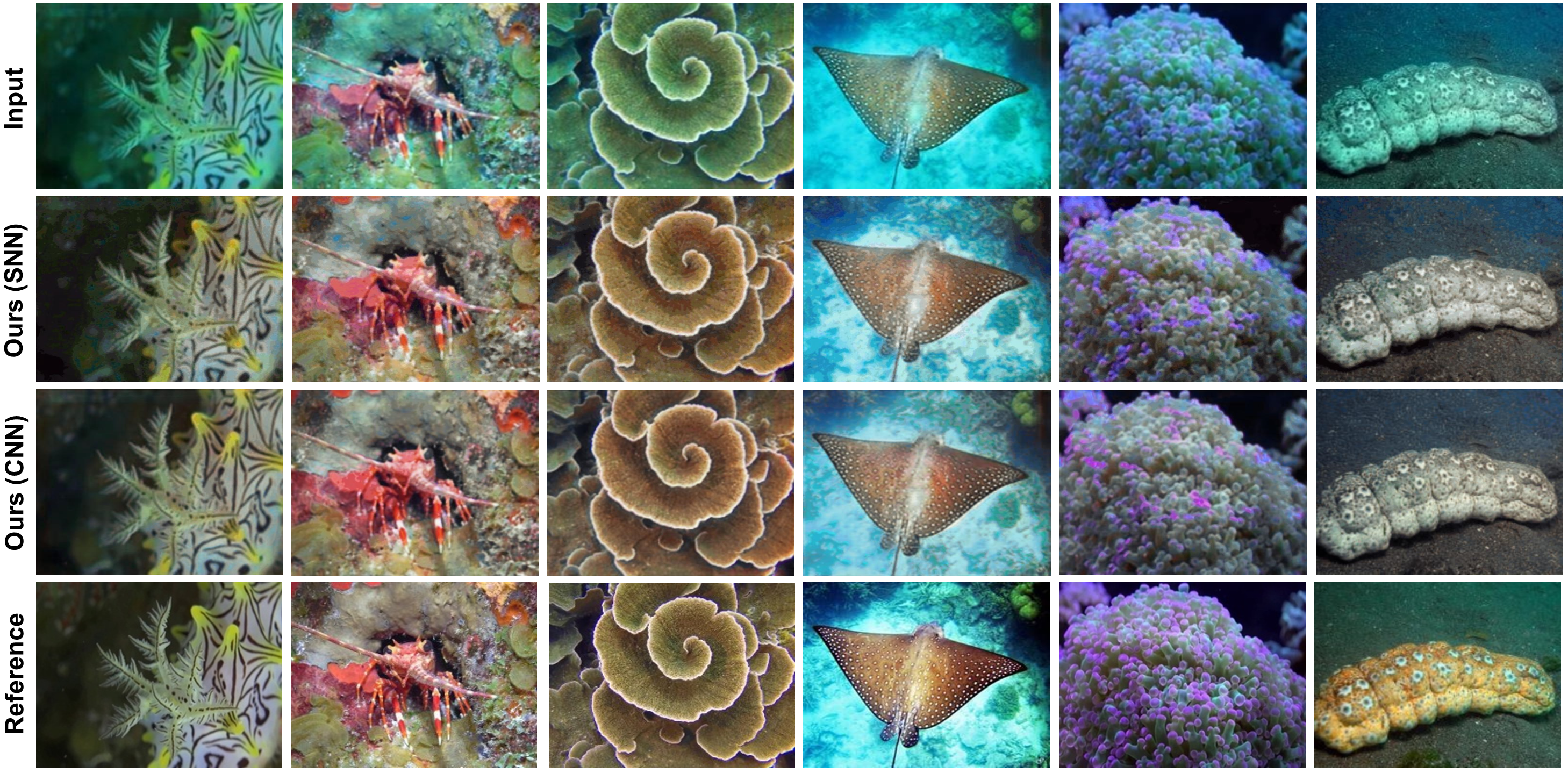}
\centering
\caption{Qualitative analysis of the proposed UIE-SNN framework for visibility enhancement on EUVP test samples.}
\label{fig:euvp_eval}
\end{figure}

The algorithmic metrics indicates that the SNNs are competitive with CNNs. In terms of hardware-oriented metrics, SNNs demonstrate a significant advantage in energy efficiency. The GFLOPs, a measure of computational complexity, are lower for SNNs (\(147.49 ~GSOPs\) on UIEB and \(84.65 ~GSOPs\) on EUVP) compared to the constant \(218.88 ~GFLOPs\) for CNNs. Energy consumption is considerably lower for SNNs, with values of \(0.1327 ~J\) for UIEB and \(0.0762 ~J\) for EUVP, in contrast to the CNN's constant \(1.0068 ~J\). This leads to substantial energy reductions of \(86.82\%\) and \(92.43\%\) for the UIEB and EUVP datasets, respectively. These results highlight that while SNNs maintain competitive algorithmic performance, they offer superior energy efficiency and lower computational demands compared to CNNs, making them highly suitable for resource-constrained environments. The qualitative analysis of the proposed UIE-SNN framework evaluated on UIEB and EUVP dataset is presented in Fig.\ref{fig:uieb_eval} and Fig.\ref{fig:euvp_eval} respectively.

The total energy consumption of learning-based algorithm depends on the number of synaptic operators, memory access and element addressing as discussed in \cite{lemaire2022analytical}. In SNN-based methods, the energy efficiency is influenced by the spike rate at each layer of the network, which directly affects the memory access and element addressing. Unlike their non-spiking counterparts, SNN-based models process spike-based data representations. This leads to a significantly lower average spike rate. This reduction in spike rate reflects the sparsity of feature maps, which, in turn, minimizes the number of active computations and memory accesses. Consequently, this sparsity substantially reduces the overall energy consumption of SNN-based methods in hardware implementations. This article focuses on algorithmic evaluations of UIE-SNN, and this inherent sparsity is anticipated to produce significant energy savings when implemented on specialised neuromorphic hardware.

\subsubsection{\textbf{Effect of Training Dataset on Energy Consumption}}\label{sec:data_energy_effect}
\begin{figure}[!h]
\includegraphics[width=\textwidth]{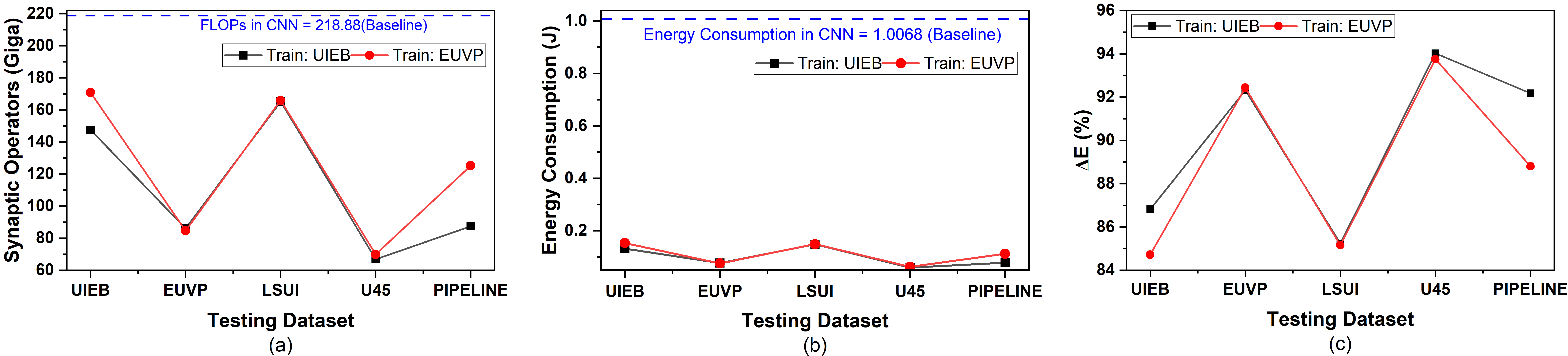}
\centering
\caption{Effect of training dataset on energy consumption. The network is trained on real underwater (UIEB) and synthetic (EUVP) datasets and  cross-validated on others: (a) Synaptic operators, (b) Energy consumption, and (c) \(\Delta E (\%)\). The blue line indicates the values computed for the baseline non-spiking counterpart.}
\label{fig:data_energy}
\end{figure}

\begin{figure}[!h]
\includegraphics[width=\textwidth]{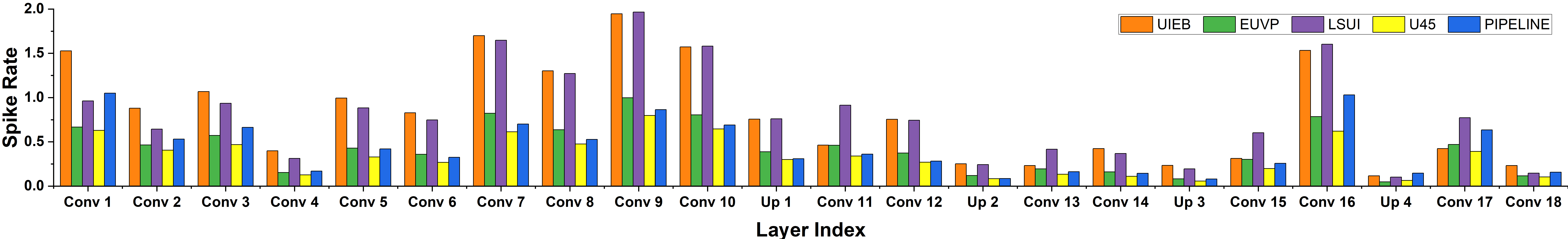}
\centering
\caption{Spike rate across each neuronal layer when trained on real underwater dataset (UIEB) and tested on others.}
\label{fig:sr_data}
\end{figure}

Unlike CNN-based frameworks, the energy consumption in SNN-based frameworks is uniquely dependent on the training dataset \cite{kim2022beyond}. In this section, we examine the generalization ability of the proposed UIE-SNN framework. The framework, trained on the UIEB and EUVP datasets, is tested using other datasets such as LSUI, U45, and a custom pipeline dataset. The UIEB, LSUI, and U45 are real-world datasets, while the EUVP and custom pipeline datasets are synthetic. Since U45 is a non-reference dataset, PSNR and SSIM values are not reported. The evaluation findings are summarized in Table \ref{table:performance_quantity}. The effect of the training dataset on energy consumption is presented in Fig.\ref{fig:data_energy} with the corresponding spike rate across each neuronal layer when trained on the real underwater dataset (UIEB) and tested on others is provided in Fig.\ref{fig:sr_data}. In all cases, the framework trained on real-world datasets provides the highest \(\Delta E (\%)\), as it effectively learns natural and complex feature representations during training, allowing the framework to enhance the visibility of underwater image sequences with lower energy consumption.

Based on everyday human experiences, it's clear that being exposed to more realistic and challenging problems before encountering unforeseen challenges can enable quicker and more efficient responses to similar situations later. For example, in mathematics exams, students who practice a broad range of complex questions can solve problems more quickly with less effort and without cognitive overload. In contrast, students who have only practiced simple problems may require more time and energy during exams. This concept is associated with cognitive load theory \cite{fischer2023using}. This difference can be attributed to the underlying neuron dynamics involved in sensory signal interpretation and memory development in the human brain. 

Analogously, suppose an ANN is trained on more realistic and challenging examples. In that case, it should be capable of extracting more discriminative features, leading to faster and more energy-efficient processing of unseen examples during inference. Although conventional ANNs mimic the human neural system, their energy consumption is determined by factors such as model architecture, the number of neurons, and other parameters, rather than the training data. The use of these kinds of neuron models limits biological plausibility. In contrast, LIF neuron dynamics used in the proposed UIE-SNN framework can validate the hypothesis that well-represented training data depicting realistic haze scenarios can reduce energy consumption during the inference stage while maintaining comparable algorithmic performance to CNNs as presented in Fig.\ref{fig:data_energy}. This represents a significant advancement in robotics applications where reducing energy consumption is a critical criterion alongside performance accuracy.

\subsection{Ablation Studies}
\label{sec:ablation}
For the effective training of LIF neurons, optimal selection of the neuron threshold and timesteps is critical, as these parameters significantly impact the temporal dynamics of neuronal behavior. The selection criteria for these parameters are detailed below. This parameter selection is based on training, validation, and testing using the UIEB dataset under the previously mentioned settings.

\subsubsection{\textbf{Effect of Threshold Selection}}
\begin{figure}[!h]
\includegraphics[width=\textwidth]{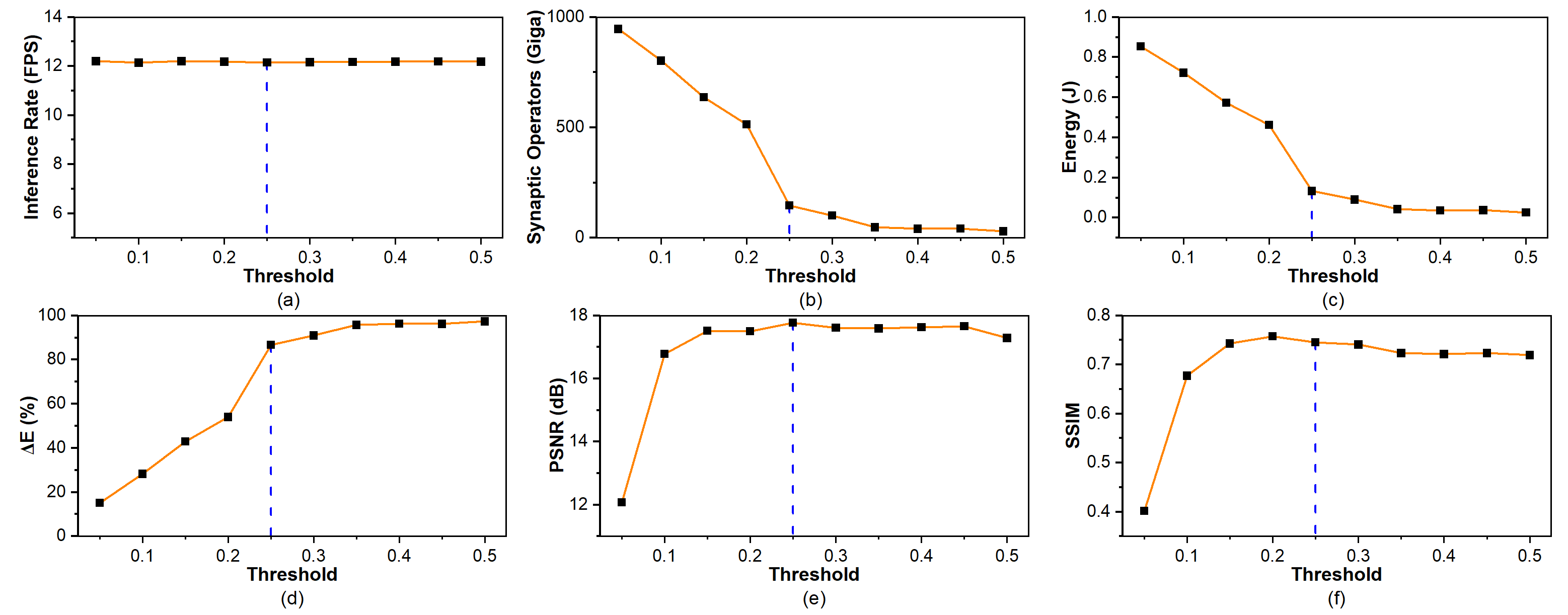}
\centering
\caption{Quantitative analysis in terms of (a) inference rate, (b) synaptic operators, (c) Energy consumption, (d) \(\Delta E (\%)\), (e) PSNR, and (f) SSIM at different threshold settings.}
\label{fig:metric_th}
\end{figure}

\begin{figure}[!h]
\includegraphics[width=\textwidth]{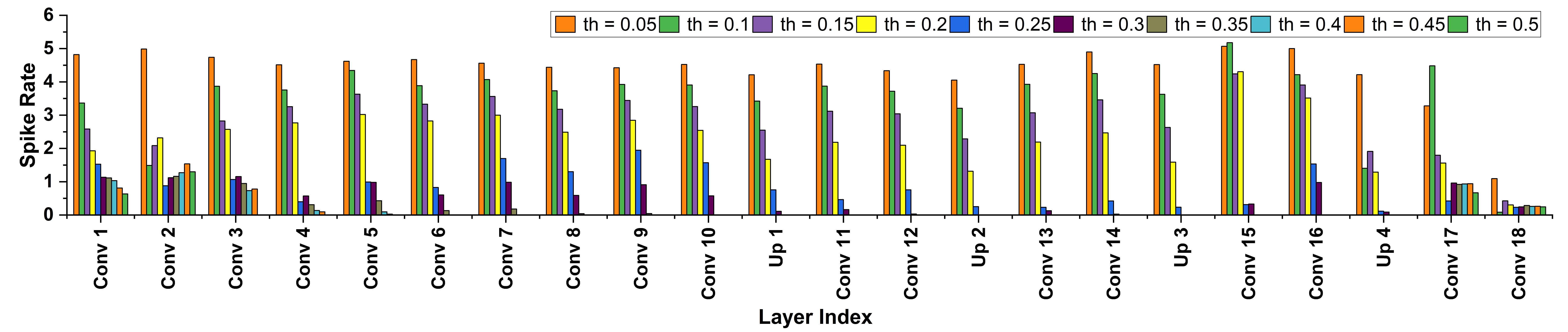}
\centering
\caption{Spike rate across each neuronal layer at different threshold settings.}
\label{fig:sr_th}
\end{figure}

To determine the optimal threshold value for each LIF neuron, an empirical analysis was conducted by fixing timestep as \(5\), setting depth of encoder-decoder as \(4\), making the membrane potential decay rate learnable, and selecting threshold from set \(\left \{ 0.05, 0.1, 0.15, 0.2, 0.25, 0.3, 0.35,0.4, 0.45, 0.5 \right \}\). The optimal threshold was selected based on algorithmic and hardware-oriented evaluation metrics presented in Fig. \ref{fig:metric_th}. The spike rate for each layer is shown in Fig.\ref{fig:sr_th}, with the last layer's spiking activity omitted as we only considered the membrane potential, leading to no reset mechanism and hence no spike firing in the last LIF neuron.

Based on the empirical results demonstrated in Fig. \ref{fig:metric_th}, the proposed UIE-SNN framework achieves an optimal trade-off between various evaluation metrics at a threshold of \(0.25\). Specifically, the inference rate remains stable across threshold values, ensuring that the processing speed is not compromised (Fig. \ref{fig:metric_th}(a)). Meanwhile, the number of synaptic operations decreases significantly up to the \(0.25\) threshold, beyond which reductions are less pronounced (Fig. \ref{fig:metric_th}(b)). This suggests that \(0.25\) is a critical point where the network operates efficiently with minimized computational complexity. Energy consumption, illustrated in Fig. \ref{fig:metric_th}(c) follows a similar trend, with a drastic decrease observed up to the 0.25 threshold, after which further reductions are marginal. This indicates that the selected threshold ensures energy-efficient operation without sacrificing performance. Additionally, the percentage of energy reduction, \(\Delta E (\%)\) continues to increase as the threshold increases, but the rate of improvement diminishes after 0.25, supporting this choice as a balance between accuracy and efficiency (Fig. \ref{fig:metric_th}(d)).

Furthermore, image quality metrics, including PSNR ans SSIM peak around the \(0.25\) threshold (Fig. \ref{fig:metric_th}(e) and \ref{fig:metric_th}(f)). This suggests that the selected threshold maximizes image quality, which is the primary objective of the proposed UIE-SNN framework. The spike rate across different layers, shown in Fig.\ref{fig:sr_th}, decreases consistently as the threshold increases, with a noticeable reduction at \(0.25\), indicating efficient neural activity that contributes to both energy savings and computational efficiency. From a theoretical perspective, the \(0.25\) threshold represents an equilibrium point where the dynamics of LIF neurons align with the goals of the network. In SNNs, the firing threshold controls the frequency of neuron spikes, influencing the sparsity of neural activity. A lower threshold results in more frequent firing, which can increase computational load and energy consumption while potentially introducing noise. Conversely, a higher threshold reduces firing rates, which might lead to the loss of critical signal information.

At a threshold of \(0.25\), the network achieves a balance between sufficient neural activity to maintain high-quality image enhancement and reduced computational and energy costs. This balance is theoretically sound, as it aligns with the inherent trade-offs in LIF neuron dynamics, where the goal is to optimize neural efficiency without compromising on the quality of the output. For further analysis, the threshold value of \(0.25\) is chosen based on its impact on various performance metrics. As the threshold increases, synaptic operations and energy consumption decrease, which is beneficial for computational efficiency. However, setting the threshold too high can lead to a significant drop in PSNR and SSIM, as seen in the figures, which could impair the network’s ability to generalize across diverse underwater scenes. By selecting a threshold of 0.25, the network maintains a balance between reducing computational load and preserving image quality, which is crucial for its performance on unseen datasets.

\subsubsection{\textbf{Effect of Timestep Selection}}
\begin{figure}[!ht]
\includegraphics[width=\textwidth]{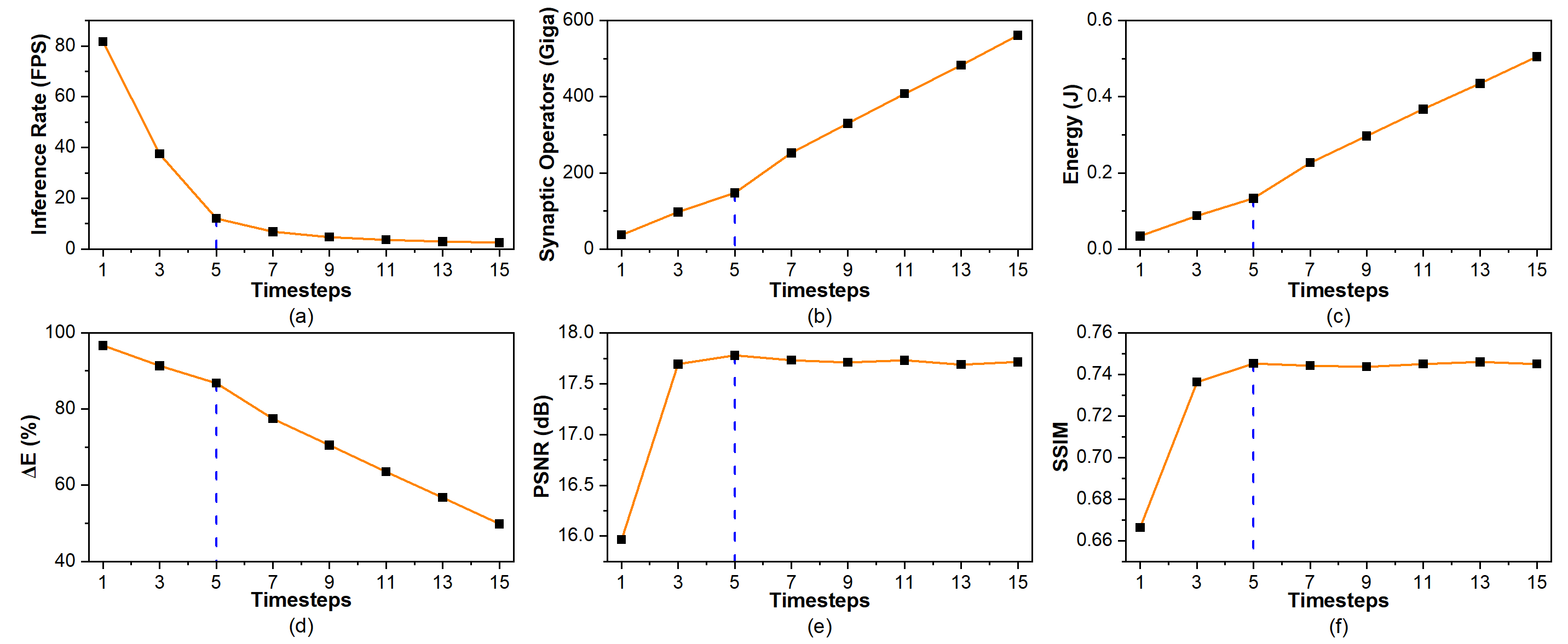}
\centering
\caption{Quantitative analysis in terms of (a) inference rate, (b) synaptic operators, (c) Energy consumption, (d) \(\Delta E (\%)\), (e) PSNR, and (f) SSIM at different timesteps settings}
\label{fig:metric_time}
\end{figure}

The number of timesteps influences the network's energy efficiency, as fewer timesteps can lead to sparser activations and reduced energy consumption. Proper tuning of timesteps balances computational load, learning convergence, and task performance, and enhances the biological plausibility of the model by simulating the temporal processing observed in real neural systems. To select the optimal number of timesteps, we analyzed model performance with eight different timesteps \(\left \{ 1, 3, 5, 7, 9, 11, 13, 15 \right \}\) keeping the threshold at \(0.25\) and layer depth at \(4\). The evaluation is  based on PSNR, SSIM, inference rate, GFLOPs, energy consumption and \(\Delta E (\%)\) (Fig.\ref{fig:metric_time}).  

The inference rate drops sharply beyond \(5\) timesteps (Fig. \ref{fig:metric_time}(a)), suggesting that further increasing the number of timesteps leads to diminishing performance in processing speed. Simultaneously, the computational cost, measured in synaptic operations, continues to rise linearly with additional timesteps, as shown in Fig. \ref{fig:metric_time}(b), indicating a higher computational overhead without a proportional gain in performance. The energy consumption depicted in Fig. \ref{fig:metric_time}(c) increases linearly with timesteps, underscoring the need to limit the number of timesteps to avoid excessive energy use. The \(\Delta E (\%)\) shows significantly large drop beyond timestep of \(5\) indicating that additional timesteps offer minimal benefits. The PSNR and SSIM metrics (Fig. \ref{fig:metric_time}(e) and \ref{fig:metric_time}(f)) demonstrate that image quality metrics plateau around timestep \(5\), with little to no improvement beyond this point. This plateau suggests that timestep \(5\) is the optimal choice for maintaining high image quality while minimizing computational cost and energy consumption. 

\begin{figure}[h!]
    \centering
    \includegraphics[width=\textwidth]{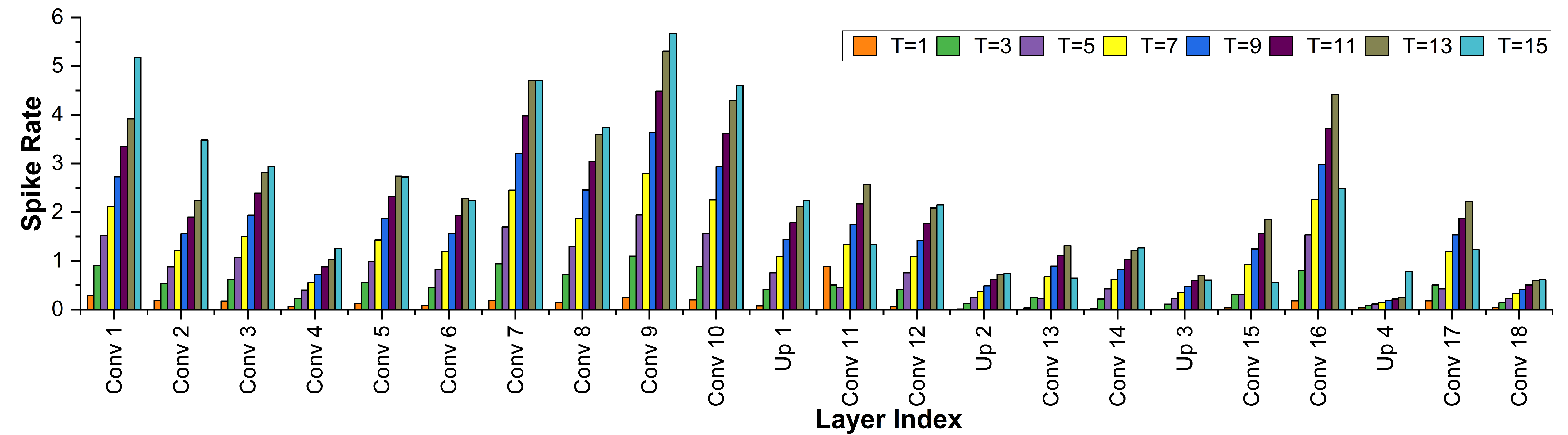}
    \caption{Spike rate across each neuronal layer at different timesteps settings.}
    \label{fig:sr_time}
\end{figure}

The spike rates across various layers at various timesteps are presented in Fig.\ref{fig:sr_time}. The spike rate tends to saturate around timestep \(5\) in most CLs, indicating that further increases in timesteps do not contribute significantly to feature representation but do increase the computational burden. This saturation reinforces the conclusion that timestep \(5\) offers an effective balance, ensuring that the neural network processes temporal dynamics adequately without unnecessary spiking activity, which would otherwise lead to increased energy and processing demands.

From a theoretical perspective, the choice of timestep \(5\) aligns with the principles of temporal coding in SNN. Temporal dynamics are essential for accurate information processing in SNNs, and timestep \(5\) appears to capture these dynamics effectively while maintaining computational efficiency. A smaller number of timesteps would likely fail to capture sufficient temporal information, while too many timesteps would result in redundant spike activity, as evidenced by the empirical data. Thus, timestep \(5\) represents an optimal point in the trade-off between accuracy and efficiency, ensuring that the network's performance remains robust without incurring unnecessary computational costs, enhancing its ability to generalize to unseen data.

\subsubsection{\textbf{Effect of Architectural Depth Selection}}
To determine the optimal number of encoder-decoder layers in the proposed architecture, an empirical analysis was conducted using a set of layer depths \(\left \{ 3, 4, 5 \right \}\), with the threshold fixed at \(0.25\) and the number of timesteps set to \(5\). The optimal depth for the proposed UIE-SNN framework was identified based on the results in Table \ref{table:metric_depth}. The results demonstrate that a 4-layer depth provides a balanced configuration, optimizing key metrics such as PSNR and SSIM while maintaining manageable levels of computational cost, measured in GFLOPs for CNN and GSOPs in SNN, and energy consumption. Specifically, the \(4\)-layer model achieves a PSNR of \(17.7801 ~dB\) and an SSIM of \(0.7454\), with computational demands of \(147.49 ~GSOPs\) and \(0.1327 ~J\) of energy. 

\begin{table}[!ht]
\renewcommand{\arraystretch}{1.3} 
\centering
\caption{Quantitative analysis for layer depth settings of \(\left \{ 3, 4, 5 \right \}\).}
\label{table:metric_depth}
\adjustbox{width=\textwidth}{\begin{tabular}{c|c|c|cc|cc|c}
\hline
\multirow{2}{*}{\textbf{Layer Depth}} & \multirow{2}{*}{\textbf{PSNR (dB) ($\uparrow$)}} & \multirow{2}{*}{\textbf{SSIM ($\uparrow$)}} & \multicolumn{2}{c|}{\textbf{GSOPs ($\downarrow$)}} & \multicolumn{2}{c|}{\textbf{Energy (J) ($\downarrow$)}} & \multirow{2}{*}{\textbf{\(\Delta E (\%)\) ($\uparrow$)}}\\
 &  &  & \textbf{\textit{CNN}} & \textbf{\textit{SNN}} & \textbf{\textit{CNN}} & \textbf{\textit{SNN}} & \\
\hline \hline
3 & 17.5473 & 0.7590 & 166.79 & 115.86 & 0.7672 & 0.1042 & 86.41\\
4 & 17.7801 & 0.7454 & 218.88 & 147.49 & 1.0068 & 0.1327 & \textbf{86.82}\\
5 & 17.6868 & 0.7452 & 270.95 & 243.45 & 1.2463 & 0.2191 & 82.42 \\
\hline
\end{tabular}}
\end{table}

A moderate network depth, such as the \(4\)-layer configuration, is crucial for effectively capturing the hierarchical nature of feature extraction required in complex tasks like UIE. Shallower architectures may fail to fully exploit these features, leading to sub-optimal performance, while deeper architectures risk overfitting and introduce significant computational overhead, as evidenced by the steep rise in GSOPs and energy consumption observed with the \(5\)-layer model. On the other hand, a shallower architecture (3 layers) reduces computational load but at the cost of lower image quality, indicating that the network may struggle with complex, unseen data. The 4-layer configuration thus offers a balanced depth that ensures robust feature extraction while avoiding overfitting, supporting the network’s ability to generalize effectively to new environments.

\section{Discussions}\label{sec:discussions}

\begin{table*}[!ht]
\renewcommand{\arraystretch}{1.3} 
\centering
\caption{Comparison of the proposed UIE-SNN framework with the SOTA SNN-based methods for pixel-wise prediction tasks.}
\label{table:snn_in_imgproc}
\adjustbox{width=\textwidth}{\begin{tabular}{c|p{0.25\linewidth}|p{0.13\linewidth}|p{0.22\linewidth}|p{0.18\linewidth}|p{0.15\linewidth}}
\hline
\textbf{Year} & \textbf{Approach} & \centering \textbf{Neuron Type} & \textbf{Training Strategy} & \textbf{Functionality} & \textbf{Dataset} \\ 
\hline \hline
2019 & Spiking Autoencoder with three fully connected layers \cite{roy2019synthesizing} & \centering LIF & Surrogate gradient with BPTT & Image reconstruction & MNIST, FMNIST \\ 

2021 & Spiking Autoencoder \cite{comcsa2021spiking} & \centering - & Surrogate gradient with BPTT & Image reconstruction & MNIST, FMNIST \\ 

2023 & DnSNN with 17 fixed dimension layers \cite{castagnetti2023spiden} & \centering IF with soft reset & Surrogate gradient with BPTT & Gaussian image denoising & Custom \\ 

2024 & Spiking Unet architecture \cite{li2024deep} & \centering IF with multi-threshold & ANN-SNN Conversion with ASF-BP based fine-tuning & Image segmentation and Gaussian image denoising & BSD68, CBSD68 \\ 

2024 & Spiking Unet with Diffusion Probabilistic Models \cite{cao2024spiking} & \centering LIF & Training-free method using Threshold Guidance & Gaussian image denoising & CelebA, CIFAR-10 \\ 

2024 & Efficient Spiking Deraining Network [ESDNet] \cite{song2024learning} & \centering LIF & Surrogate gradient with BPTT & Deraining & Rain12, Rain200L \\ 
\hline
\textbf{2024} & \textbf{UIE-SNN [19-layered deep convolutional SNN framework] [Ours]} & \centering \textbf{LIF with learnable decay} & \textbf{Surrogate gradient with BPTT} & \textbf{Underwater Image Enhancement} & \textbf{UIEB, EUVP, LSUI, U45, PIPELINE} \\
\hline
\end{tabular}}
\end{table*}

An overview of SNN-based frameworks applied to pixel-wise prediction tasks is presented in Table \ref{table:snn_in_imgproc}. From the early implementations of spiking autoencoders for image reconstruction on MNIST and FMNIST datasets in 2019 and 2021, to more complex models like DnSNN with dimension layers and Spiking Unet architectures for tasks like Gaussian image denoising and segmentation, the field has shown significant progress. Notably, the 2023 models introduced multi-threshold IF neurons and ANN-SNN conversion techniques to enhance performance on challenging datasets such as BSD68 and CBSD68. In 2024, the introduction of SNN frameworks integrated with Diffusion Probabilistic Models and efficient spiking dehazing networks further diversified the applications, including tasks like Gaussian image denoising and deraining. These frameworks are trained and tested on low-resolution images, making them unsuitable for directly processing high-resolution, complex real-world data, such as images captured in underwater environments. The quality of these underwater images is significantly degraded due to factors like turbidity, color cast, and distortion.

Our proposed model, a 19-layer deep convolutional spiking framework, stands out due to its unique combination of LIF neurons with learnable decay and the use of surrogate gradient-based backpropagation (BPTT) through time strategy for training. This model is tailored for visibility enhancement of high resolution (\(512\)x\(512\)) underwater image sequences, a niche yet critical application, as evidenced by its robust performance across diverse datasets including UIEB, EUVP, LSUI, U45, and custom PIPELINE datasets. The novelty of our approach lies in its ability to generalize well across both real-world and synthetic datasets while maintaining high energy efficiency and computational feasibility, thereby addressing a significant gap in the current literature where most models are used for denoising low-resolution images. This makes our proposed UIE-SNN framework not only relevant, but also a pioneering solution in the realm of UIE tasks using SNNs.

\begin{table}[h!]
\renewcommand{\arraystretch}{1.3} 
\centering
\caption{Comparison of the proposed UIE-SNN framework with the SOTA analytical-based UIE methods.}
\label{table:sota_conv}
\adjustbox{width=0.7\textwidth}{\begin{tabular}{c|c c|c c}
\hline
\multirow{3}{*}{\textbf{Method}} & \multicolumn{2}{c|}{\textbf{Dataset: UIEB}}& \multicolumn{2}{c}{\textbf{Dataset: EUVP}} \\
 & \multicolumn{2}{c|}{(Real-World)} & \multicolumn{2}{c}{Synthetic)} \\
 & \textbf{\textit{PSNR}} & \textbf{\textit{SSIM}} & \textbf{\textit{PSNR}} & \textbf{\textit{SSIM}} \\
\hline \hline
DCP \hfill(2010) \cite{he2010single}  & 14.2300 & 0.7037 & 17.5800 & 0.7473 \\
UDCP \hfill(2013) \cite{drews2013transmission} & 11.9286 & 0.6441 & 14.4190 & 0.8140 \\
Retinex \hfill(2014) \cite{fu2014retinex} & 17.6800 & \textbf{0.8199} & 15.9600 & 0.7213  \\
IBLA \hfill(2017) \cite{peng2017underwater} & 15.5610 & 0.7390 & 16.9223 & \textbf{0.8643}  \\
ULAP \hfill(2018) \cite{song2018rapid} & 16.3200 & 0.7578 & 19.6200 & 0.7964  \\
\hline 
\textbf{UIE-SNN \hfill{[Ours]}}  & \textbf{17.7801} & 0.7454 & \textbf{23.1725} & 0.7890 \\
\hline
\end{tabular}}
\end{table}

The comparison between the proposed UIE-SNN framework and the SOTA analytical-based and learning-based methods is depicted in Table \ref{table:sota_conv} and \ref{table:sota_learn}, respectively. The learning-based methods are categorized into CNN, GAN, Attention, Transformer, and finally, the SNN-based methods. The proposed UIE-SNN offers several distinct advantages over both analytical and \(2^{nd}\) generation learning-based UIE methods. One of the most significant advantages of the proposed framework is its remarkable energy efficiency. Our model consumes just \(0.1327~J\), which is substantially lower than other learning-based methods. For instance, the attention-based DICAM method requires \(0.4888 ~pJ\), which is \(3.68\times\) more energy consumption with the proposed UIE-SNN despite its higher algorithmic performance in SSIM metric and its lowest number of GFLOPs. The Ucolor method, a CNN-based method, consumes \(30.77\times\) more energy than UIE-SNN, while GAN-based methods like LFT-DGAN consume \(1.2157~pJ\) (\(9.16\times\) more), and transformer-based MMIETransformer method consume \(0.7544~pJ\) (\(5.68\times\)) more. This efficiency is primarily due to the use of ACC operations instead of the traditional MAC operations employed in conventional neural networks. The energy required to perform a MAC operation is approximately \(4.6 ~pJ\), whereas an ACC operation consumes only \(0.9~pJ\). This significant reduction in energy consumption makes the SNN framework particularly suitable for deployment in power-constrained environments, such as underwater robotics or edge devices, where energy is a critical resource.

\begin{table}[!ht]
\centering
\caption{Comparison of the proposed UIE-SNN framework with the SOTA learning-based UIE methods.}
\label{table:sota_learn}
\begin{threeparttable}
\renewcommand{\arraystretch}{1.3} 
\adjustbox{width=\textwidth}{
\begin{tabular}{c|cc|cc|c|c|c}
\hline
\multirow{3}{*}{\textbf{Method}} & \multicolumn{2}{c|}{\textbf{Dataset: UIEB}}& \multicolumn{2}{c|}{\textbf{Dataset: EUVP}} & \textbf{Number} & \textbf{Energy} & \textbf{Energy}\\
 & \multicolumn{2}{c|}{(Real-World)}& \multicolumn{2}{c|}{(Synthetic)} & \textbf{of} & \textbf{Consumption} & \textbf{Ratio}\tnote{\(\#\)} \\
 & \textbf{\textit{PSNR}} & \textbf{\textit{SSIM}} & \textbf{\textit{PSNR}} & \textbf{\textit{SSIM}} & \textbf{Operators}\tnote{\(\star\)} &  \textbf{(J)} & \((\times)\) \\
\hline \hline
\textbf{CNN-based method:} \hfill{} &  &  &  &  &  &  &  \\
WaterNet \hfill(2020) \cite{li2020underwater}  & 21.1700 & 0.8689 & 20.4200 & 0.8711 & 285.80 & 1.3147 & 9.9 \(\times\) \\
Shallow-UWNet \hfill(2021) \cite{naik2021shallow}  & 18.2780 & 0.8550 & - & - & 216.14 & 0.9942 & 7.49\(\times\)\\
Ucolor \hfill(2021) \cite{li2021underwater}  & 18.8400 & 0.7612 & 18.8300 & 0.7713 & 887.80 & 4.0839 & 30.77\(\times\)\\
PUIE-Net \hfill(2022) \cite{fu2022uncertainty}  & 21.3800 & 0.882 & - & - & 300.04 & 1.3802 & 10.40\(\times\)\\
PDCFNet  \hfill(2024) \cite{zhang2024pdcfnet} & \textbf{27.3700} & 0.9200 & 26.6700 & 0.8400 & 172.77 & 0.7947 & 5.99\(\times\)\\
UIEConv \hfill(2024) \cite{du2024physical} & 24.1197 & 0.9283 & - & - & 243.06 & 1.1181 & 8.43\(\times\)\\

\textbf{GAN-based method:} \hfill{} &  &  &  &  &  &  &  \\
PUGAN \hfill(2023) \cite{cong2023pugan} & 19.9700 & 0.7848 & 21.0500 & 0.8641 & 144.18 & 0.6632 & 4.99\(\times\)\\
DGD-cGAN \hfill(2024) \cite{gonzalez2024dgd} & - & - & 24.0860 & 0.7560 & 188.88 & 0.8688 & 6.55\(\times\)\\
LFT-DGAN \hfill(2024) \cite{zheng2024learnable} & 24.4591 & 0.8284 & - & - & 264.28 & 1.2157 & 9.16\(\times\)\\

\textbf{Attention-based method:} \hfill{} &  &  &  &  &  &  &  \\
LANet \hfill(2022) \cite{liu2022adaptive}  & 24.0600  & 0.9085 & 25.8200 & 0.8668 & 355.37 & 1.6347 & 12.32\(\times\)\\
Deep-WaveNet \hfill(2023) \cite{sharma2023wavelength}  & 23.3000 & 0.9199 & 24.6700 & 0.8953 & 145.18 & 0.6678 & 5.03\(\times\)\\
DICAM \hfill(2024) \cite{tolie2024dicam} & 24.4300 & \textbf{0.9375} & 25.1300 & \textbf{0.9131} & \textbf{106.26} & 0.4888 & 3.68\(\times\)\\  

\textbf{Transformer-based method:} \hfill{} &  &  &  &  &  &  &  \\
U-Trans \hfill(2023) \cite{peng2023u} & 22.9100 & 0.9100 & - & - & 132.04 & 0.6074 & 4.56\(\times\)\\
MFMN  \hfill(2024) \cite{zheng2024multi} & 24.2067 & 0.8992 & 23.3908 & 0.8896 & 205.61 & 0.9458 & 7.13\(\times\)\\
ICDT-XL/2  \hfill(2024) \cite{nie2024image} & - & - & 28.1673 & 0.8744 & 237.28 & 1.0915 & 8.22\(\times\)\\
CFPNet \hfill(2024) \cite{fu2024cfpnet} & 25.6900 & 0.8860 & - & - & 111.84 & 0.5145 & 3.88\(\times\)\\
MMIETransformer \hfill(2024) \cite{kulkarni2024multi} & 23.2800 & 0.9100 & \textbf{30.1400} & 0.8740 & 164.00 & 0.7544 & 5.68\(\times\)\\
\hline
\textbf{SNN-based method:} \hfill{} &  &  &  &  &  &  &  \\
\textbf{UIE-SNN \hfill{[Ours]}} & 17.7801 & 0.7454 & 23.1725 & 0.7890 & 147.49 & \textbf{0.1327} & \textbf{1\(\times\)} \\ 
\hline
\end{tabular}
}
\begin{tablenotes}
    \item [\(\star\)] Number of operators are measured in GSOPs for SNN-based method \\
    and in GFLOPs for others.
    \item [\(\#\)] The energy ratio is calculated as \(\frac{E_{method}}{E_{UIE-SNN}}\).  
\end{tablenotes}
\end{threeparttable}
\end{table}

In addition to its energy efficiency, the proposed UIE-SNN framework also demonstrates competitive performance metrics, particularly in terms of PSNR and SSIM values. Despite its simpler architecture, which relies solely on spiking convolution operations, the SNN framework achieves PSNR and SSIM values that are comparable to those of more complex and resource-intensive models like Ucolor and LANet. For instance, the proposed model achieves a PSNR of \(17.7801 ~dB\) and an SSIM of \(0.7454\) on the UIEB dataset and a PSNR of \(23.1725 ~dB\) and an SSIM of \(0.7890\) on the EUVP dataset. These results highlight the potential of spiking-based models to deliver high-quality performance with much lower computational and energy demands. 

When evaluated over the conventional analytical-based UIE methods, the proposed UIE-SNN framework significantly outperforms them, further emphasizing its strengths in both performance and energy efficiency. On the UIEB (real-world) dataset, the SNN framework achieves a PSNR of \(17.7801 ~dB\) and an SSIM of \(0.7454\), which are higher than those obtained by traditional methods like DCP (\(14.2300 ~dB\), \(0.7037\) SSIM) and IBLA (\(15.5610 ~dB\), \(0.7390\) SSIM). This demonstrates that the SNN framework is not only capable of producing clearer and more accurate images but also excels in preserving structural details that are often lost in conventional methods. On the EUVP (synthetic) dataset, the performance gap becomes even more pronounced. The proposed UIE-SNN framework achieves a PSNR of \(23.1725 ~dB\) and an SSIM of \(0.7890\), outperforming all other analytical-based methods by a considerable margin. For example, the Retinex method, which is one of the more advanced conventional techniques, achieves a PSNR of \(15.9600 ~dB\) and an SSIM of \(0.7213\), significantly lower than those of the SNN framework. This substantial difference highlights the ability of the SNN framework to handle synthetic data effectively. The ability of our model to handle both real-world and synthetic datasets with superior performance demonstrates its robustness and versatility. Conventional analytical-based methods, such as DCP and IBLA, often rely on heuristic approaches that are specifically tailored to certain types of image degradation or environmental conditions. As a result, their performance can degrade significantly when applied to images outside of their designed scenarios. In contrast, the SNN framework, with its learning-based approach, adapts to a wide range of conditions, resulting in consistently high performance across different datasets.

However, the proposed UIE-SNN framework does face certain limitations. One notable limitation is the performance gap when compared to the highest-performing learning-based UIE methods. While the SNN framework achieves comparable results, it falls short of the performance metrics obtained by some of the most advanced models, which achieve higher PSNR and SSIM values due to the incorporation of advanced techniques like GAN and transformers with attention mechanisms like the MMIETransformer method, which achieves \(30.1400 ~dB\) PSNR on EUVP dataset. This limitation is largely due to the simpler architecture of our UIE-SNN framework, which, while advantageous for energy efficiency, limits its capacity to capture the complex features that more sophisticated architectures can. 

Despite these limitations, the proposed UIE-SNN framework presents significant potential for real-time applications, particularly in power-constrained environments. Its low energy consumption and asynchronous spike-based information flow make it an ideal candidate for deployment in scenarios where traditional neural networks may be too resource-intensive. The difference between \(0.1327 ~J\) and \(0.4888 ~J\) of energy consumption during model inference is substantial, particularly when considered over the entire duration of a mission. In such scenarios, the ability to minimize energy usage can directly impact the robot's operational lifespan, mission duration, and overall effectiveness. In the context of real-time deployment, where the system must process data continuously, the model consuming \(0.1327 ~J\) presents a clear advantage. The lower energy consumption directly translates to prolonged battery life, which is crucial for underwater robots that often operate in remote or inaccessible locations where recharging or battery replacement is not feasible. Additionally, in an environment where every subsystem, including propulsion, sensors, and communication, is drawing from a limited power source, reducing the energy required for computation allows for a more efficient allocation of resources, potentially extending the robot’s operational time.

Edge devices, which are typically designed to operate with limited power resources, benefit significantly from energy-efficient models. In such devices, even small differences in energy consumption can accumulate, leading to substantial effects on battery life, especially in systems that perform frequent inferences. Furthermore, the difference in energy consumption of the proposed UIE-SNN framework and other existing learning-based models can impact the overall scalability of deploying such models across a fleet of robots or multiple edge devices. In large-scale deployments, the cumulative energy savings from using a more efficient model can translate into significant operational cost reductions, both in terms of energy expenditure and maintenance requirements. This can be particularly advantageous when numerous devices are deployed simultaneously, each requiring efficient energy management to maximize their collective performance. The lower energy consumption model not only extends operational time and reduces the thermal load but also provides a scalable solution for large deployments. Unless the higher-energy model offers substantial improvements in performance that are critical to the application, the energy savings and operational benefits of the lower-energy model make it the more practical choice for real-time deployment in these environments.

\section{Conclusion and Future Research Guidelines}
This article presents a novel approach for improving the visibility of underwater image sequences through the utilization of direct training of a convolutional SNN-based framework. The proposed framework demonstrates a significant reduction in energy consumption while ensuring competitive algorithmic performance. By leveraging accumulation operations, the framework achieves notable energy savings, making it ideal for power-constrained environments like underwater robotics. The article explored the dependency of training data representation on the energy reduction capacity of the models with the same configuration. The model trained on complex real-world data representation is able to reduce $85\%$ of energy consumption with its non-spiking counterpart.

Moving forward, future research endeavors could focus on integrating advanced architectural components, such as attention mechanisms, into the SNN framework to enhance its feature-capturing capabilities while maintaining energy efficiency. Exploring hybrid models that combine SNNs with conventional deep learning techniques could also offer a balanced approach to high-performance, energy-efficient computing. Furthermore, real-time implementation and deployment of the SNN framework in practical scenarios, as well as expanding its application beyond UIE to other domains, could establish SNNs as a pivotal technology in developing next-generation, low-power computing systems.

\section*{Acknowledgments}
The authors acknowledge the project “Heterogeneous Swarm of Underwater Autonomous Vehicles”, a collaborative research project between the Technology Innovation Institute (Abu Dhabi)  and Khalifa University  (contract no. TII/ARRC/2047/2020). This work is also supported by Khalifa University under Awards No. RC1-2018-KUCARS-8474000136. 

We express our gratitude to Dr. Rachmad Vidya Wicaksana Putra, Dr. Alberto Marchisio and Prof. Muhammad Shafique (eBrain Lab, Division of Engineering, New York University-Abu Dhabi, UAE) for their valuable feedback and insightful suggestions on Section \ref{perform_eval}) of the article.

\section*{Code Availability}
The code will be made publicly available upon publication of the manuscript.

\section*{Supplementary Materials}
All results associated with the performance evaluation of the proposed model, including detailed energy and synaptic operator calculations, can be accessed at  \href{https://www.dropbox.com/scl/fo/l3wwb1to6m8aj7lgz3hfb/AKqSWswyVVeAFzbNVe7Un9U?rlkey=qxdpi9mpcc4rc34i3b1fdlk1e&st=2ss1qnul&dl=0}{UIE-SNN:Results}.


\bibliographystyle{elsarticle-num} 
\bibliography{UIE_SNN-Revised_Manuscript}

\end{document}